\documentclass[floatfix,
aps,
10pt,
twocolumn,
superscriptaddress,
amsmath,amssymb,
physrev,
]{revtex4-2}

\usepackage{graphicx}
\usepackage{dcolumn}
\usepackage{bm}
\usepackage{physics}
\usepackage{color}

\newcommand{\Am}[0]{\hat{a}}
\newcommand{\Ap}[0]{\hat{a}^\dagger}
\newcommand{\Qa}[0]{\mathrm{Q_a}}
\newcommand{\Qb}[0]{\mathrm{Q_b}}
\newcommand{\Qc}[0]{\mathrm{Q_c}}
\newcommand{\fgge}[0]{{gfeg}}
\newcommand{\fggetxt}[0]{\ket{gf} \text{--} \ket{eg}} 
\newcommand{\fggetxtbc}[0]{\ket{gf}_\mathrm{bc} \text{--} \ket{eg}_\mathrm{bc}} 
\usepackage[colorlinks=true,citecolor=blue,linktocpage=true,linkcolor=blue,filecolor=blue,urlcolor=magenta]{hyperref}
\begin{document}

\preprint{APS/123-QED}

\title{
High-fidelity all-microwave CZ gate with partial erasure-error detection
\\via a transmon coupler
}

\author{Shotaro Shirai}
\email{\text{mailto:}shotaro.shirai@riken.jp}
\affiliation{RIKEN Center for Quantum Computing (RQC), Wako, Saitama 351--0198, Japan}
\affiliation{Komaba Institute for Science (KIS), The University of Tokyo, Meguro-ku, Tokyo, 153-8902, Japan}

\author{Shinichi Inoue}
\affiliation{Department of Applied Physics, Graduate School of Engineering, The University of Tokyo, Bunkyo-ku, Tokyo 113-8656, Japan}

\author{Shuhei Tamate}
\affiliation{RIKEN Center for Quantum Computing (RQC), Wako, Saitama 351--0198, Japan}

\author{Rui Li}
\affiliation{RIKEN Center for Quantum Computing (RQC), Wako, Saitama 351--0198, Japan}

\author{Yasunobu Nakamura}
\affiliation{RIKEN Center for Quantum Computing (RQC), Wako, Saitama 351--0198, Japan}
\affiliation{Department of Applied Physics, Graduate School of Engineering, The University of Tokyo, Bunkyo-ku, Tokyo 113-8656, Japan}

\author{Atsushi Noguchi}
\email{\text{mailto:}u-atsushi@g.ecc.u-tokyo.ac.jp}
\affiliation{Komaba Institute for Science (KIS), The University of Tokyo, Meguro-ku, Tokyo, 153-8902, Japan}
\affiliation{RIKEN Center for Quantum Computing (RQC), Wako, Saitama 351--0198, Japan}
\affiliation{Inamori Research Institute for Science (InaRIS), Kyoto-shi, Kyoto 600-8411, Japan}

\date{\today}

\begin{abstract}
Entangling gates between neighboring physical qubits are essential for quantum error correction. Implementing them in an all-microwave manner simplifies signal routing and control apparatus of superconducting quantum processors. We propose and experimentally demonstrate an all-microwave controlled-Z (CZ) gate that achieves high fidelity while suppressing residual ZZ interactions. Our approach utilizes a fixed-frequency transmon coupler and multi-path coupling, thereby sufficiently reducing the net transverse interaction between data transmons to suppress residual ZZ interactions. The controlled phase arises from the dispersive frequency shift of the $\fggetxt$ transition between the coupler and one of the data transmons conditioned on the state of the other data transmon. Driving the transitions at the midpoint of two dispersively shifted resonance frequencies induces state-dependent geometric phases to achieve the CZ gate. Crucially, with this scheme, we can maintain a small net transverse interaction between two data transmons while increasing the coupling between the data and coupler transmons to accelerate the CZ-gate speed. Additionally, we measure the coupler state after the gate to detect a subset of decoherence-induced failures that occur during the gate operation. These events constitute erasure errors with known locations, enabling erasure-aware quantum error-correcting codes to improve future logical qubit performance.
\end{abstract}

\maketitle


\section{Introduction}
Progress toward large-scale quantum computers has been driven by the discovery of quantum algorithms that promise computational speedups~\cite{doi:10.1137/S0097539795293172, Nielsen_Chuang_2010, RevModPhys.86.153} and has been accelerated by advancements in hardware engineering. However, quantum states are fragile and susceptible to environmental disturbances, making it challenging to retain information and execute operations reliably. Consequently, quantum error correction (QEC) is essential. Efforts toward QEC span across multiple platforms, including superconducting circuits~\cite{10.1063/1.5089550, RevModPhys.93.025005}, trapped ions~\cite{10.1063/1.5088164, pino2021demonstration}, neutral atoms~\cite{Bluvstein2024, bluvstein2025architecturalmechanismsuniversalfaulttolerant}, photonics~\cite{10.1063/1.5115814, Alexander2025}, and other solid-state systems~\cite{RevModPhys.95.025003, Takeda2022}. Recently, experiments have shown a trend of decreasing logical memory errors as the number of physical qubits increases~\cite{Acharya2025, bluvstein2025architecturalmechanismsuniversalfaulttolerant}, indicating a shift from experimental validation to logical performance evaluation of QEC codes.

In current QEC experiments, two-qubit gate errors account for a significant portion of the error budget~\cite{Acharya2025}, motivating intensive efforts to improve the fidelity of the two-qubit gate. Notably, approaches based on transmons and magnetic flux-tunable couplers have achieved high-fidelity two-qubit gates~\cite{YanTC, PhysRevLett.125.120504, PhysRevLett.127.080505, RuiDTC, Chen2025} utilizing their high on–off coupling ratios. However, magnetic flux-tunable couplers require additional flux-bias lines, which increase sensitivity to magnetic-field fluctuations and introduce extra decoherence channels. An alternative approach is to employ all-microwave gates with fixed-frequency transmons, which reduce control wiring complexity and support longer coherence times. Several all-microwave gate schemes have been developed for fixed-frequency transmons using the cross-resonance~(CR) interaction, either resonantly or off-resonantly~\cite{CR_2010Rigetti, ProcedureCR, JAZZ1, KandalaCR, RotaryECR, siZZle}, four-wave-mixing parametric interaction~\cite{bSWAP_2012Poletto, Krinner_CZ}, and dispersive interactions mediated by a resonator coupler~\cite{RIP_2016Paik, kumph2024demonstrationripgatesquantum}. Both CR and four-wave-mixing parametric gates face an inherent trade-off between the gate speed and residual ZZ interaction: increasing the effective transverse interaction to speed up gates typically increases residual ZZ interaction, degrading single- and two-qubit gate performances. To mitigate this trade-off, solutions such as utilizing additional coupling elements to use multi-path coupling~\cite{KandalaCR, SS} and combining superconducting qubits with opposite anharmonicities~\cite{JAZZ1, fasciati2024complementingtransmonintegratinggeometric} have been introduced. The multi-path coupling approach, especially for the CR gate, is practical only in the so-called straddling regime, i.e., within the limited anharmonicity. The latter requires hybrid systems, such as flux qubits coupled to transmons, which demand different fabrication processes. Another scheme is the resonator-induced phase~(RIP) gate~\cite{RIP_2016Paik, kumph2024demonstrationripgatesquantum}. This approach utilizes a coupler resonator and avoids the trade-off by employing dispersive interactions between the coupler resonator and data qubits coupled to it, rather than the net transverse interaction between the data qubits. When a microwave pulse drives the coupler resonator, state-dependent dispersive shifts produce four distinct time-evolution paths in its phase space conditioned on the four computational bases. This characteristic, however, makes it difficult to return the resonator to its initial state in order to prevent state leakage. As a result, necessities for adiabatic pulse envelopes and larger detuning of the drive frequency from the resonator frequency set a limit on the gate speed~\cite{RIP_opt}.

In this work, we propose and experimentally demonstrate an all-microwave controlled-Z (CZ) gate employing a fixed-frequency transmon coupler and the concept of the RIP gate. This scheme overcomes the trade-off and operates outside the straddling regime. Our circuit design also incorporates the idea of multi-path coupling~\cite{SS, KandalaCR, kumph2024demonstrationripgatesquantum, fors2024comprehensiveexplanationzzcoupling}, which enables substantial suppression of the residual ZZ interaction, without compromising gate speed. To implement the CZ gate, we drive the fixed-frequency transmon coupler to induce a four-wave-mixing parametric transition between one data transmon and the coupler transmon. The transition occurs between the coupler's $\ket{f}$ state and the $\ket{e}$ state of one data transmon, and its resonance frequency acquires a shift that depends on the state of the other data transmon. This state-dependent frequency shift generates a controlled phase within the computational subspace, enabling a CZ-gate fidelity of $99.7(1)\%$ with a gate time of $140$~ns. We refer to this coupler-induced CZ gate as the Transmon-Induced Phase~(TIP) gate. In contrast to the RIP gate, the parametric Rabi oscillation between the coupler and the data transmon follows only two time-evolution paths, conditioned on the states of the other data transmon qubit. As a result, we can implement the CZ gate with a simple pulse shape without the need for adiabatic constraints or composite pulses to suppress coupler leakage.

Even with a perfect pulse, state leakage in the coupler can occur due to decoherence. However, the TIP gate features an intrinsic error detection mechanism that enables the diagnosis of gate failures. The TIP gate intentionally populates the coupler to the second excited state and is supposed to depopulate it back to the ground state. The gate failure can be detected by measuring the excitation of the coupler at the end. We use randomized benchmarking to quantify the fraction of detectable two-qubit gate error via post-gate measurement of the coupler, and estimate that approximately $45(4)\%$ of two-qubit Clifford-gate error, including CZ-gate error events, are detected. These detection events can be regarded as erasure errors at the known location. Combining them with erasure-aware decoders improves quantum error-correction performance~\cite{Goto2009, Wu2022, KubicaErasure} and suppresses the generation and propagation of nonlocal errors due to coupler leakage~\cite{Varbanov2020, Miao2023}.

This paper is organized as follows. In Sec.~\ref{sec:gate_scheme}, we describe the gate scheme proposed in this work. Section~\ref{sec:dev_cal} provides details of the experimental device and explains the method for suppressing the residual ZZ interaction using multi-path coupling. In Sec.~\ref{sec:cz_impl}, we present the calibration procedure for the CZ gate and summarize the results of its performance evaluation. Section~\ref{sec:ped_demo} reports the method for detecting qubit gate errors through measurements of the coupler transmon, together with quantitative performance evaluation. Finally, in Sec.~\ref {sec:conc_disc}, we discuss potential improvements to the proposed approach and conclude the paper.

\section{\label{sec:gate_scheme}Transmon-induced phase gate}
\begin{figure}
    \centering
    \includegraphics{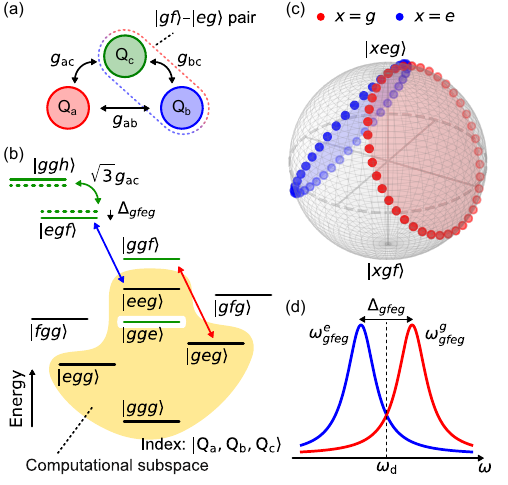}
    \caption{Transmon-induced phase~(TIP) gate. (a)~Schematic of the system. Two data transmons Q$_\mathrm{a}$ and Q$_\mathrm{b}$ interact transversely with the coupler transmon Q$_\mathrm{c}$ with the strengths $g_\mathrm{ac}$ and $g_\mathrm{bc}$, respectively. The direct interaction strength between the data qubits is $g_\mathrm{ab}$.
    (b)~Energy-level diagram of the three-transmon system. The computational subspace is shaded yellow, and levels involving excitation of the coupler transmon are shaded green. Red and blue arrows indicate the $\fggetxt$ transitions between Q$_\mathrm{b}$ and Q$_\mathrm{c}$ conditioned on the state of Q$_\mathrm{a}$. Green dashed lines indicate the Q$_\mathrm{a}$-dependent frequency shift of the $\fggetxt$ transition, $\Delta_{gfeg}$.
    (c)~Bloch-sphere depiction of the ideal evolution of the $\fggetxt$ transitions in subspaces labeled by the state of Q$_\mathrm{a}$. The red trajectory corresponds to the case where $\Qa$ is in $\ket{g}$, and the blue trajectory is for $\Qa$ in $\ket{e}$.
    (d)~Schematic spectra of the $\fggetxt$ transition. When Q$_\mathrm{a}$ is in $\ket{e}$, the $\fggetxt$ transitions frequency shifts downward in the case of $\omega_{ggh}>\omega_{egf}$. In the ideal case, the drive frequency $\omega_\mathrm{d}$ is set to the midpoint between the two resonance peaks.
    }
    \label{fig:1}
\end{figure}
We consider the system with three transmons (labeled by $\Qa$, $\Qb$, and $\Qc$) as shown in Fig.\,\ref{fig:1}(a). The total Hamiltonian is modeled as coupled Duffing oscillators under the rotating-wave approximation,
\begin{align}
    \hat{H}/\hbar &= \sum_i \left(\omega_i\Ap_i\Am_i + \frac{\alpha_i}{2}\Ap_i\Ap_i\Am_i\Am_i\right) \notag\\
    &+ \sum_{i\neq j} g_{ij}(\Ap_i \Am_j + \Am_i\Ap_j),
\label{eq:H}
\end{align}
where $\hbar$ is the reduced Planck constant; $\omega_i$ and $\alpha_i$ denote the fundamental frequency and anharmonicity of each transmon Q$_i$~($i \in \{\mathrm{a, b, c}\}$); $\Am_i$ and $\Ap_i$ are the annihilation and creation operators; and $g_{ij}$ is the transverse interaction strength between Q$_i$ and Q$_j$. We assume the dispersive regime $|g_{ij}/\Delta_{ij}|\ll 1$, where $\Delta_{ij}=\omega_i-\omega_j$. We truncate each transmon to the third excited state in the following analysis. Here we consider the case $\omega_\mathrm{a} < \omega_\mathrm{b} < \omega_\mathrm{c}$. In this ordering, the drive frequency to activate the TIP gate lies above the coupler frequency, which avoids many undesired parametric transitions below the fundamental frequencies of the transmons that are induced by the negative anharmonicity of the transmon. The state of the three-transmon system is represented in the order $\ket{\Qa, \Qb, \Qc}$. When explicitly specifying particular transmons, we use subscripts to indicate them, such as $\ket{\mathrm{Q}_i}_i$ for a single transmon or $\ket{\mathrm{Q}_i, \mathrm{Q}_j}_{ij}$ for a pair of transmons. 

We first focus on the pair Q$_\mathrm{b}$–Q$_\mathrm{c}$. When a microwave drive is applied to the coupler, modeled as
\begin{align}
    \hat{H}_\mathrm{d}/\hbar = \Omega_\mathrm{d} \cos{\omega_\mathrm{d} t} \left(\Ap_\mathrm{c} + \Am_\mathrm{c} \right),
\label{eq:Hdrive}
\end{align}
we can parametrically induce the $\fggetxt$ transition by tuning the drive frequency to be the energy difference of $\ket{gf}_\mathrm{bc}$ and $\ket{eg}_\mathrm{bc}$ states, as shown in several studies~\cite{Krinner_CZ, fogi_int}. Here, $\omega_\mathrm{d}$ and $\Omega_\mathrm{d}$ are the drive frequency and amplitude, respectively. The Rabi-oscillation frequencies associated with the $\fggetxt$ parametric transition can be expressed perturbatively as~\cite{Krinner_CZ, fogi_int}
\begin{align}
    \Omega_{\fgge}^e \approx \Omega_{\fgge}^g \approx   \frac{\sqrt{2}g_\mathrm{bc}\alpha_\mathrm{c}\Omega_\mathrm{d}}{\Delta_\mathrm{bc}(\Delta_\mathrm{bc}+\alpha_\mathrm{c})}.
    \label{eq:fgge}
\end{align}
The superscripts indicate the state of Q$_\mathrm{a}$, which is not part of the driven pair. Note that, within second-order perturbation theory, the Rabi-oscillation frequencies conditioned on the state of Q$_\mathrm{a}$ are identical. In the present system, Q$_\mathrm{a}$ couples to Q$_\mathrm{c}$, and the resulting interaction between $\ket{egf}$ and $\ket{ggh}$ mainly produces a state-dependent frequency shift $\Delta_{\fgge}$ of the $\fggetxt$ transition frequency~\cite{MAP_2013Chow, SFQ_gate}, as depicted in Fig.\,\ref{fig:1}(b). The frequency shift, which depends on the state of Q$_\mathrm{a}$, is perturbatively given by
\begin{align}
    \Delta_{\fgge} &= \omega_\fgge^g - \omega_\fgge^e \notag\\
    &\approx -\frac{g_{\mathrm{ac}}^2}{\Delta_\mathrm{ac}} + \frac{2g_{\mathrm{ac}}^2}{\Delta_\mathrm{ac}+\alpha_\mathrm{c}} \notag\\
    &+ \frac{3g_{\mathrm{ac}}^2}{\Delta_\mathrm{ac}+2\alpha_\mathrm{c}} - \frac{4g_{\mathrm{ac}}^2}{\Delta_\mathrm{ac}-\alpha_\mathrm{a}+\alpha_\mathrm{c}}, \label{eq:fgge_shift}
\end{align}
where $\omega_\fgge^g$ and $\omega_\fgge^e$ are the corresponding transition frequencies for Q$_\mathrm{a}$ in $\ket{g}_\mathrm{a}$ and $\ket{e}_\mathrm{a}$. These expressions are obtained using second-order perturbation theory, neglecting $g_\mathrm{ab}$ for simplicity (see Appendix~\ref{sec: Perturbation} for details).
We now restrict the discussion to the $\ket{kgf}$--$\ket{keg}$ subspace. For each state $k \in \{g, e\}$ of Q$_\mathrm{a}$, the effective Hamiltonian, in the frame rotating at $\omega_\mathrm{d}$, is given as follows:
\begin{align}
    \hat{H}_\fgge^k/\hbar &= -\frac{\delta^k}{2}\hat{Z}_\fgge + \frac{\Omega_\fgge^k}{2}\hat{X}_\fgge.
    \label{eq: H_fgge}
\end{align}
Here, $\hat{Z}_\fgge=\dyad{gf}{gf}_\mathrm{bc}-\dyad{eg}{eg}_\mathrm{bc}$, $\hat{X}_\fgge=\dyad{eg}{gf}_\mathrm{bc}-\dyad{gf}{eg}_\mathrm{bc}$, and $\delta^k = \omega_\fgge^k - \omega_\mathrm{d}$. After a full cycle generalized Rabi oscillation [Fig.\,\ref{fig:1}(c)], the initial state $\ket{keg}$ acquires the geometric phase given as follows~\cite{Leek_GP}:
\begin{align}
    \phi_{keg} = \pi\left(1 - \frac{\delta^k}{{\Omega'}^k}\right), \label{eq:gp_keg}
\end{align}
where ${\Omega'}^k = \sqrt{{\Omega_\fgge^k}^2 + {\delta^k}^2}$ is the generalized Rabi-oscillation frequency. Ignoring the difference of the $\fggetxt$ Rabi-oscillation frequencies, the leakage-free conditions are
\begin{align}
    {\Omega'}^g &= {\Omega'}^e =: \Omega', \label{eq:leak_free_c0}\\
    \Omega' t_\mathrm{g} &= 2\pi, \label{eq:leak_free_c1}
\end{align}
where $t_\mathrm{g}$ is the gate time of the CZ gate. Eqs.\,\eqref{eq:leak_free_c0} and \eqref{eq:leak_free_c1} lead that the drive frequency should be $\omega_\mathrm{d} = (\omega_\fgge^g + \omega_\fgge^e)/2$ as shown in Fig.\,\ref{fig:1}(d).
Assuming that the phases of the other computational bases remain unchanged, the CZ gate condition is~\cite{GotoDTC}
\begin{align}
    - \phi_{geg} + \phi_{eeg} = \pi. \label{eq:CZ_condition}
\end{align}
Using Eqs.\,\eqref{eq:gp_keg} and \eqref{eq:CZ_condition}, the generalized Rabi-oscillation frequency should satisfy $\Omega' = \Delta_\fgge$. Under these conditions, we obtain the leakage-free CZ gate. In summary, the required detuning, gate time, and $\fggetxt$ Rabi-oscillation frequencies are $\delta^{g} = -\delta^{e} = \Delta_\fgge/2$, $t_\mathrm{g} = 2\pi/\Delta_\fgge$, and $\Omega_\fgge^g = \Omega_\fgge^e = \sqrt{3}\Delta_\fgge/2$, respectively. In contrast to the RIP gate, which involves four state-dependent paths, the TIP gate involves only two, allowing for the analytical determination of the optimal detuning and gate duration even in the case that the $\fggetxt$ Rabi-oscillation frequencies are different due to higher-order effects than second-order perturbation. To account for this, we introduce the ratio $r=\Omega_\fgge^e/\Omega_\fgge^g$. Imposing $\pi$ geometric phase and equal generalized Rabi-oscillation frequency in both subspaces yields
\begin{align}
    \frac{\delta'}{\Delta_{\fgge}} = \frac{-1 + \sqrt{r^2(r^2-1)+1}}{r^2 - 1},
\label{eq: gen_ratio}
\end{align}
where $\delta' = \omega_\fgge^g - \omega_\mathrm{d}$. This relation ensures that suitable pulse parameters exist to implement the CZ gate while minimizing leakage by selecting the drive frequency appropriately. Similar analyses are found in the following references~\cite{FastLogic, SFQ_gate}.

\section{\label{sec:dev_cal}Device setup and characterization}
\begin{figure}
    \centering
    \includegraphics{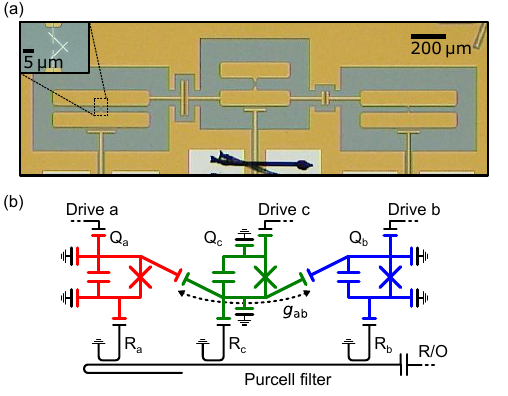}
    \caption{Device structure. (a)~Optical micrograph of the fabricated three-transmon device (inset: Al/AlO$_x$/Al Josephson junction). Most features (yellow) use domain-matched epitaxially grown titanium-nitride (TiN) electrodes~\cite{DMTiN} on a Si substrate (gray). (b) Equivalent circuit of the coupled-transmon system. Q$_\mathrm{a}$, Q$_\mathrm{b}$, and Q$_\mathrm{c}$ denote two data transmons and a coupler transmon, respectively. Each transmon has its own drive line. R$_\mathrm{a}$, R$_\mathrm{b}$, and R$_\mathrm{c}$ denote the readout resonators, which connect with a common Purcell filter for multiplexed readout. 
    }
    \label{fig: design}
\end{figure}
In this work, we use a circuit consisting of three capacitively coupled transmons [Figs.\,\ref{fig: design}(a) and (b)]. Each transmon couples to its own $\lambda/4$ coplanar-waveguide (CPW) readout resonator, and the three readout resonators share a common Purcell filter. We estimate the device parameters listed in Table~\ref{tab:device_params} by fitting spectroscopic data with the model Hamiltonian in Eq.\,\eqref{eq:H}. Appendices~\ref{sec: coherence} and \ref{sec: Circ} summarize the coherence-time measurements and the detailed design guidelines for this device, respectively.

\begin{table}
    \caption{Estimated device parameters and coherence times. Values in parentheses denote the $\pm1\sigma$ standard error.}
    \centering
    \begin{ruledtabular}
    \begin{tabular}{l@{\hspace{1em}}c@{\hspace{1em}}c@{\hspace{1em}}c}
                             & Q$_\mathrm{a}$ & Q$_\mathrm{b}$ & Q$_\mathrm{c}$ \\ \hline
    $\omega_{i}/2\pi$ (MHz) & $4464$ & $4985$ & $5746$ \\
    $\alpha_i/2\pi$ (MHz)      & $-225$ & $-230$ & $-314$ \\
    $g_\mathrm{ac}/2\pi$ (MHz)      & $63$ & & \\
    $g_\mathrm{bc}/2\pi$ (MHz)      & & $31$ & \\
    $g_\mathrm{ab}/2\pi$ (MHz)      & $3$ & & \\
    $T_{1}$ (µs)       & $168(32)$ & $80(23)$ & $42(5)$ \\
    $T_{1f}$ (µs)       &  &  & $32(5)$ \\
    $T_{2}$ (µs)  & $90(11)$ & $85(10)$ & $64(10)$ \\
    $T_{2f}$ (µs)  &  &  & $26(4)$ \\
    \end{tabular}
    \end{ruledtabular}
    \label{tab:device_params}
\end{table}

\subsection{Intrinsic static-ZZ suppression}
In fixed-frequency transmon systems, it is important to design circuits that minimise the residual ZZ interaction. To second order in perturbation theory, the residual ZZ interaction is given by~\cite{ProcedureCR}
\begin{align}
    \xi_\mathrm{zz} \approx \frac{2g_\mathrm{eff}^2(\alpha_\mathrm{a} + \alpha_\mathrm{b})}{(\Delta_\mathrm{ab}-\alpha_\mathrm{a})(\Delta_\mathrm{ab}+\alpha_\mathrm{b})}.
\label{eq:zz}
\end{align}
Here, $g_\mathrm{eff}$ is the net transverse interaction between transmons $\Qa$ and $\Qb$. Assuming that the coupler remains in the ground state and applying the rotating-wave approximation, $g_\mathrm{eff}$ takes the second-order form~\cite{YanTC},
\begin{align}
    g_\mathrm{eff} \approx g_\mathrm{ab} + \frac{g_\mathrm{ac}g_\mathrm{bc}}{2}\left( \frac{1}{\Delta_\mathrm{ac}} + \frac{1}{\Delta_\mathrm{bc}} \right).
\label{eq:g_eff}
\end{align}
Hence, engineering $g_\mathrm{eff}$ enables suppression of the residual ZZ interaction. Prior works achieved this with resonator couplers or tunable couplers~\cite{KandalaCR, SeteCoupler}. In contrast, our device uses the electrode layout shown in Fig.\,\ref{fig: design}(a): one electrode of the coupler forms a series‑capacitor path between $\Qa$ and $\Qb$ as shown in Fig.\,\ref{fig: design}(b) as the dotted arrow, contributing to the first term $g_\mathrm{ab}$ in Eq.\,\eqref{eq:g_eff}. With this layout, all transverse interactions have positive signs, and the second term in Eq.\,\eqref{eq:g_eff} is overall negative since the coupler has the highest frequency. Therefore, the first and second terms in Eq.\,\eqref{eq:g_eff} partially cancel each other, which suppresses the residual ZZ interaction without adding extra coupling elements through Eq.\,\eqref{eq:zz}. In addition, as discussed in the previous section, the TIP-gate speed scales as $2\pi/\Delta_\fgge$. Therefore, decreasing $g_\mathrm{eff}$ does not slow the gate.

In this work, we design the sample aiming at a parameter regime that yields CZ gate time of 50--150~ns with required drive amplitude of 50--200~MHz. These values are achievable when the coupling-to-detuning ratios satisfy $|g_{i\mathrm{c}}/\Delta_{i\mathrm{c}}|\in[0.05,0.06]$ for each data–coupler pair (see Appendix~\ref{sec: Circ}). In this regime, the magnitudes of both terms in Eq.\,\eqref{eq:g_eff} lie in the sub-MHz to few-MHz range, allowing efficient reduction of $g_\mathrm{eff}$ and suppression of residual ZZ interaction without intentionally adjusting the direct transverse interaction $g_\mathrm{ab}$. Note that the resulting sample has slightly smaller coupling-to-detuning ratios than the targets, which does not affect the demonstration of the TIP gate. Appendix~\ref{sec: Circ} offers a detailed parameter analysis and an example of circuit layout in which this suppression does not occur.

\subsection{Static-ZZ interaction measurement}
\begin{figure}
    \centering
    \includegraphics{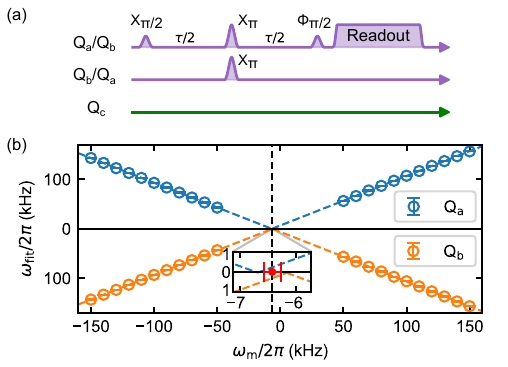}
    \caption{Characterization of residual interaction. (a) Joint amplification of ZZ (JAZZ) pulse sequence that measures the ZZ interaction between the data transmons~\cite{JAZZ0, JAZZ1}. The sequence is repeated with the roles of the measured data transmons exchanged. To improve the fitting accuracy, the phase of the second half-$\pi$ pulse~($\Phi_{\pi/2}$) is swept with the idling time $\tau$ such that $\phi/\tau = \omega_\mathrm{m}$. (b) Oscillation frequencies extracted from the JAZZ sequence. The frequencies are obtained by fitting the experimental data to a decaying cosine. The upper and lower panels show the results for $\mathrm{Q_a}$ and $\mathrm{Q_b}$, respectively. Note that the vertical axis for Q$_\mathrm{b}$ is inverted in the bottom half. Dashed lines indicate linear fits to the positive- and negative-slope branches. The residual ZZ interaction, averaged over $\mathrm{Q_a}$ and $\mathrm{Q_b}$, is indicated by the black dashed line and by the red dot in the inset.
    }
    \label{fig:zz_meas_jazz}
\end{figure}
In the experiment, we measure the residual ZZ interaction using the pulse sequence in Fig.\,\ref{fig:zz_meas_jazz}(a), referred to as the joint-amplification-of-ZZ (JAZZ) sequence~\cite{JAZZ0, JAZZ1}. In this sequence, we map the ZZ-induced phase accumulation between the target and control qubits onto the final state of the target qubit. To cancel local phases, we apply $\pi$ pulses to both qubits at the midpoint of the sequence. The target's ground-state population then exhibits a cosine oscillation due to the residual ZZ interaction. To improve the accuracy of the fit, we vary the measurement axis by sweeping the phase $\phi$ of the final $\pi/2$ pulse along with the waiting time $\tau$ as $\phi = \omega_\mathrm{m} \tau$. Neglecting decoherence, the final ground-state probability oscillates as $\cos[(\omega_\mathrm{m} + \xi_\mathrm{zz})t]$. By sweeping $\omega_\mathrm{m}$ and linearly fitting the extracted oscillation frequency versus $\omega_\mathrm{m}$, the intersection point yields an estimate of the residual ZZ interaction $\xi_\mathrm{zz}$. We repeat the measurement with the roles of target and control swapped and summarize the results in Fig.\,\ref{fig:zz_meas_jazz}(b). Averaging the two outcomes, we estimate the residual ZZ interaction between data transmons Q$_\mathrm{a}$ and Q$_\mathrm{b}$ to be $6.4(2)$~kHz.
In addition, we perform the single-qubit standard randomized benchmarking~(SRB) and simultaneous RB, both of which yield single-qubit gate fidelities exceeding $99.9\%$. The absence of significant fidelity degradation during simultaneous RB confirms that the effect of residual ZZ interaction on single-qubit gates is negligible.

\section{\label{sec:cz_impl}CZ-gate implementation}
\subsection{Calibration}
\begin{figure*}
    \centering
    \includegraphics{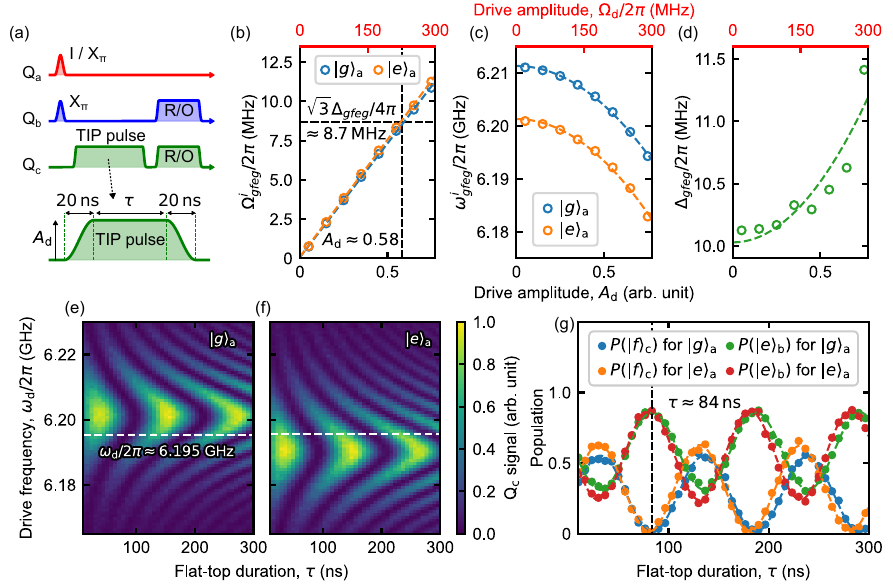}
    \caption{Implementation of TIP gate. (a)~Pulse sequence used to measure the $\fggetxt$ Rabi oscillations between Q$_\mathrm{b}$ and Q$_\mathrm{c}$. To condition on the state of Q$_\mathrm{a}$, we use two variants of the Q$_\mathrm{a}$ sequence, with and without an X$_\pi$ pulse. A flat-top microwave pulse drives the $\fggetxt$ transition, called a TIP pulse, with raised-cosine edges (edge length 20 ns).
    (b,c,d) $\fggetxt$ Rabi-oscillation frequency~$\Omega_{gfeg}^i$, resonance frequency~$\omega_{gfeg}^i$, and the state-dependent frequency shift $\Delta_{gfeg}$. For each drive amplitude $A_\mathrm{d}$, we record Rabi oscillations by sweeping both the pulse duration $\tau$ and the drive frequency $\omega_\mathrm{d}$ of the TIP pulse and fit the patterns with an exponentially decaying cosine. From the fit, we extract the minimum $\fggetxt$ Rabi-oscillation frequency for each subspace~[plotted in (b)] and the corresponding drive frequency [plotted in (c)]. Panel (d) shows $\Qa$-state-dependent frequency shift of $\fggetxt$ resonance $\Delta_{gfeg} = \omega^e_{gfeg} - \omega^g_{gfeg}$. The dashed lines in (b) and (c) indicate linear and quadratic fits, respectively. In (d), the dashed line is the difference between the fits in (c). In~(b), the horizontal black dotted line marks $\sqrt{3} \, \Delta_{gfeg}/2$ at zero drive amplitude, calculated from the fit. The vertical black dotted line marks the drive amplitude at which the fit for the case with $|g\rangle_\mathrm{a}$ intersects the horizontal line. The upper horizontal axes show the drive strength $\Omega_\mathrm{d}$, calibrated with the linear coefficient of the linear fit to the data for $|g\rangle_\mathrm{a}$ in (b).
    (e,f)~$\fggetxt$ Rabi-oscillation chevron patterns at the drive amplitude indicated by the vertical black dotted line in (b). Horizontal white dashed lines mark the condition that minimizes the $\ket{f}_\mathrm{c}$-state population simultaneously across the subspaces corresponding to $\ket{g}_\mathrm{a}$ and $\ket{e}_\mathrm{a}$ after one full cycle of the Rabi oscillation, determined from exponentially decaying-cosine fits.
    (g) Rabi oscillations for Q$_\mathrm{b}$ and Q$_\mathrm{c}$ in each subspace under the drive-frequency condition indicated by the white dashed lines in (e) and (f). The dashed curves show fits to an exponentially decaying cosine. The vertical black dotted line denotes the initial estimate of the flat-top duration $\tau$ of the TIP pulse implementing the CZ gate.}
    \label{fig:gfeg_sweep}
\end{figure*}

\begin{figure}
    \centering
    \includegraphics{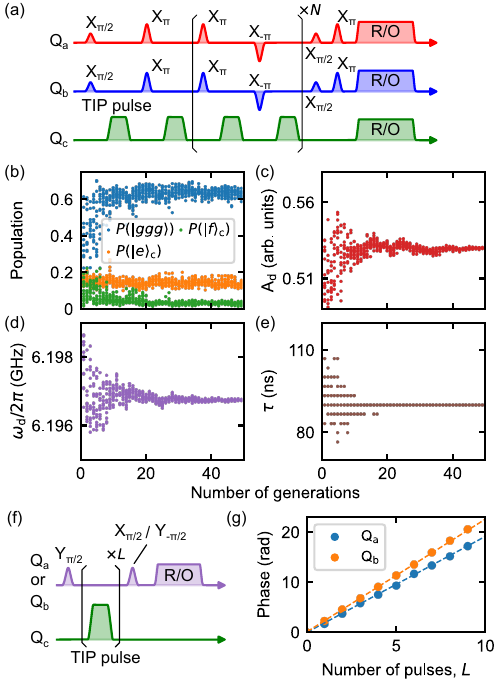}
    \caption{Optimization of TIP gate. (a)~JAZZ2-N pulse sequence. Here, we use $N = 2k$, and $k=5$. (b) Evolution of the state populations $P(|ggg\rangle)$, $P(|e\rangle_\mathrm{c})$, and $P(|f\rangle_\mathrm{c})$, corresponding respectively to the states $\ket{ggg}$, $\ket{e}_\mathrm{c}$, and $\ket{f}_\mathrm{c}$, during the optimization process using the JAZZ2-N pulse sequence. The optimization cost is $P(|ggg\rangle)$. As the optimization proceeds, $P(|f\rangle_\mathrm{c})$ decreases, indicating suppression of the leakage into $\ket{f}_\mathrm{c}$ state, while $P(|e\rangle_\mathrm{c})$ remains unchanged, confirming that population leakage into $\ket{e}_\mathrm{c}$ state is not increasing. The optimization is performed using the CMA-ES algorithm implemented in Optuna~\cite{optuna}, with a population size of 15 per generation. After 50 generations, the parameter set used for subsequent experiments is taken as the mean of the parameters from the final generation. (c,d,e) Convergence behavior of the drive amplitude~$A_\mathrm{d}$, drive frequency~$\omega_\mathrm{d}$, and flat-top pulse duration~$\tau$, during the optimization process. The flat-top duration takes discrete values due to the time-resolution granularity of the AWGs used in the experiment.
    (f) Pulse sequence to calibrate the local phases induced by the TIP pulse on each data transmon. In the calibration protocol, the number of TIP pulses, $L$, is increased, and tomography is performed along the $x$- and $y$-axes to measure the accumulated phase. (g) Measured accumulated phases induced by the TIP pulse on each data transmon. The dashed lines represent linear fits, and the slope extracted from each fit determines the virtual-Z-gate phase required for the CZ-gate implementation.}
    \label{fig:cz_cal}
\end{figure}

The CZ gate calibration begins by estimating the drive frequency, amplitude, and pulse duration. We characterize the $\fggetxt$ Rabi oscillations using the pulse sequence in Fig.\,\ref{fig:gfeg_sweep}(a). The initial state is prepared as either $\ket{geg}$ or $\ket{eeg}$, after which the TIP pulse is applied to the coupler transmon. By sweeping the drive frequency, amplitude, and pulse length, we record time-domain Rabi-oscillation patterns and fit them to an exponentially decaying sinusoidal function to extract the Rabi-oscillation frequency for each drive frequency and amplitude. The resonance frequencies of $\fggetxtbc$ transition conditioned on the states of $\Qa$ can be extracted as the drive frequencies that minimize the frequencies of Rabi oscillations. The extracted resonance frequencies are shown in Fig.\,\ref{fig:gfeg_sweep}(c), and the state-dependent frequency shift is plotted in Fig.\,\ref{fig:gfeg_sweep}(d). These frequencies provide an initial parameter regime for optimizing the CZ gate that satisfies the leakage-free condition explained in Sec.\,\ref{sec:gate_scheme}. Although Eq.\,\eqref{eq:fgge_shift} predicts no amplitude dependence, the data show a clear dependence, which we attribute to ac Stark shifts with subspace-dependent amplitude dependencies.

From Sec.~\ref{sec:gate_scheme}, the CZ gate is realized when $\Omega_\fgge=\sqrt{3} \,\Delta_\fgge/2$. In Fig.\,\ref{fig:gfeg_sweep}(b), the horizontal black dotted line marks $\sqrt{3} \,\Delta_\fgge/2$ evaluated by using the value of $\Delta_{\fgge}$ at zero drive extrapolated from Fig.\,\ref{fig:gfeg_sweep}(d). The vertical black dotted line marks the drive amplitude at which the fit for the Rabi-oscillation frequency for $\ket{g}_\mathrm{a}$ intersects this horizontal line. At this amplitude, the Rabi-oscillation frequency ratio between different states of $\Qa$ is $r\approx 1.03$, and the corresponding detuning deviation $\delta'$ from $\Delta_\fgge/2$ is about 0.2~MHz. Because this deviation is small, we neglect the difference of the $\fggetxt$ Rabi-oscillation frequency for each state of $\Qa$ and use the values obtained when $\Qa$ is in $\ket{g}$ as the reference in the following analysis.
The Rabi-oscillation patterns for each state of $\Qa$ at this intersection condition are shown in Figs.\,\ref{fig:gfeg_sweep}(e) and\,(f). The white dashed lines indicate the drive frequencies that minimize the $\ket{f}_\mathrm{c}$ population simultaneously across the subspaces after one period of the coupler-transmon $\fggetxt$ Rabi oscillation, as determined from exponentially decaying-cosine fits. Figure~\ref{fig:gfeg_sweep}(g) shows the $\ket{e}_\mathrm{b}$ and $\ket{f}_\mathrm{c}$ populations at these drive frequencies. From these fits, we obtain an initial estimate of the flat-top duration of the TIP pulse required to implement the CZ gate.

Further calibration of the controlled phase $\phi_\mathrm{cp}$ uses the JAZZ2-N pulse sequence~\cite{RuiDTC} shown in Fig.\,\ref{fig:cz_cal}(a). The final population of $\ket{gg}_\mathrm{ab}$ depends on the controlled phase given as follows: 
\begin{align}
    P({\ket{gg}_\mathrm{ab}}) = \frac{1 - \cos{((N+1)\phi_\mathrm{cp}})}{2}.\label{eq:jazz2_pop}
\end{align}
Here,  $N=2k$, and $k$ is a positive integer. Increasing $N$ enhances the phase sensitivity. To make the sequence sensitive to leakage in the coupler transmon, we choose the population of $\ket{ggg}$ as the optimization cost and maximize it via \textit{in-situ} optimization, thereby steering $\phi_\mathrm{cp}$ toward $\pi$. We perform this optimization using the CMS-ES algorithm~\cite{optuna}, which is robust to noisy cost functions. In the optimization process, each generation consists of 15 candidates. The optimizer evaluates the cost of each candidate and updates the sampling distribution to favor better candidates. We adopt the mean parameter values from the final generation after 50 iterations as the TIP-pulse parameters. The optimization traces are shown in Figs.\,\ref{fig:cz_cal}(b)–(e). The TIP-pulse amplitude, frequency, and flat-top duration converge as the optimization proceeds. Figure~\ref{fig:cz_cal}(b) also plots the populations of the coupler's $\ket{e}_\mathrm{c}$ and $\ket{f}_\mathrm{c}$ states, confirming that leakage does not increase during optimization. For this run, we use slightly different initial parameters from those in Fig.\,\ref{fig:gfeg_sweep}, which empirically yield better optimization results.

We next calibrate local phases induced by the ac Stark shift of the TIP pulse on each data transmon, using the sequence in Fig.\,\ref{fig:cz_cal}(f). After preparing each data transmon in a superposition state, we apply $N$ TIP pulses to the coupler and measure the accumulated local phase by extracting the $x$- and $y$-components of the Bloch vector via quantum state tomography. Figure~\ref{fig:cz_cal}(g) shows the representative data. The accumulated phase increases linearly with $L$, and the slope of a linear fit yields the required virtual-Z (VZ) phase $\phi_\mathrm{Z}^\mathrm{a}$ ($\phi_\mathrm{Z}^\mathrm{b}$) to realize the CZ gate with matrix diag$(1,1,1,-1)$. 

\subsection{CZ-gate benchmarking}
\begin{figure}
    \centering
    \includegraphics{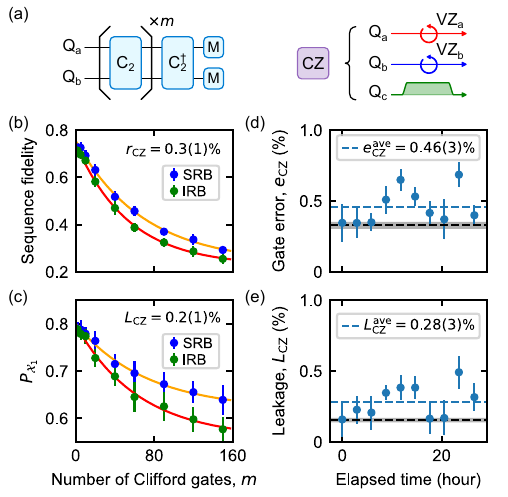}
    \caption{
    Randomized benchmarking. (a)~Gate sequence for standard randomized benchmarking (SRB) and the CZ pulse sequence. In interleaved randomized benchmarking (IRB), the CZ gate is inserted after each two-qubit Clifford gate. We average over 10 random Clifford sequences, and each sequence is repeated 1000 times.
    (b)~Sequence fidelities, $P(|gg\rangle_\mathrm{ab})$, obtained from SRB and IRB.
    (c)~Population in the computational subspace, $P_{\mathcal{X}_1}$, during SRB and IRB.
    In~(b)~and\,(c), solid curves show exponential-decay fits. Using the fitting results, the estimated gate errors and the leakage rates are $r_\mathrm{SRB}=1.08(9)\%$, $r_\mathrm{IRB}=1.38(9)\%$, $L_\mathrm{SRB}=0.52(8)\%$, and $L_\mathrm{IRB}=0.68(9)\%$, respectively. (d)~CZ-gate error, $e_\mathrm{CZ}$, measured over approximately one day. The blue dashed line indicates the mean CZ-gate error. (e)~CZ-gate leakage rate, $L_\mathrm{CZ}$, measured over approximately one day. The blue dashed line indicates the mean leakage rate. In~(d)~and\,(e), black dashed lines show coherence limits estimated using the analytical error model and independently measured coherence times, with $\pm1\sigma$ standard error represented by the gray shaded bands. The first points in both panels correspond to the estimated results in~(b) and\,(c). In (b)--(e), error bars show $\pm1\sigma$ standard errors.
    }
    \label{fig:SRB_IRB_2Q}
\end{figure}
To evaluate the performance of the calibrated CZ gate, we perform RB-type experiments. Figure~\ref{fig:SRB_IRB_2Q}(a) shows the SRB gate sequence and the pulse decomposition of the CZ gate. During each TIP pulse, we apply VZ gates simultaneously to the data qubits to compensate for the local phase shifts estimated in the calibration. As shown in Fig.\,\ref{fig:gfeg_sweep}(a), the TIP pulse has cosine-shaped 20-ns rising and falling edges, and the optimization gives the 90-ns flat-top duration. With a 10-ns buffer inserted between pulses, the total CZ gate duration is 140~ns. All single-qubit Clifford gates are implemented using two $\mathrm{X}_{\pi/2}$ gates and three VZ gates. The pulse implementing X$_{\pi/2}$ is a 40-ns cosine-shaped pulse with DRAG correction~\cite{DRAG}. Including the buffer times, the single-qubit Clifford gate duration is 100~ns. Two-qubit Clifford gates contain zero to three CZ gates plus single-qubit Clifford gates. On average, a two-qubit Clifford gate requires 1.5\,CZ gate and five single-qubit Clifford gates per two-qubit Clifford gate~\cite{Barends2014}. In interleaved RB (IRB), a CZ gate is inserted after every two-qubit Clifford gate except the final one. In SRB and IRB, we classify the $\ket{g}$, $\ket{e}$, and $\ket{f}$ states of each transmon and calculate the populations of all 27 joint states, and then calculate the sequence fidelity, $P(\ket{gg}_\mathrm{ab})$, and the computational-subspace population, $P_{\mathcal{X}_1} = P(\ket{ggg}) + P(\ket{egg}) + P(\ket{geg}) + P(\ket{eeg})$, respectively.

Figure~\ref{fig:SRB_IRB_2Q}(b) shows typical SRB and IRB results. The sequence fidelities are fit to an exponentially decaying function $A \lambda_\mathrm{SRB/IRB}^m + B$, where $\lambda_\mathrm{SRB/IRB}$ is the decay parameter and $A$ and $B$ account for state-preparation-and-measurement (SPAM) errors. From $\lambda_\mathrm{SRB/IRB}$ we calculate the Clifford-gate error as
\begin{align}
    r_\mathrm{SRB/IRB} = \frac{d-1}{d} \left(1 - \lambda_\mathrm{SRB/IRB} \right), \label{eq:r_srb}
\end{align}
where $d = 2^2$ is the dimension of the computational subspace. As shown in Fig.\,\ref{fig:SRB_IRB_2Q}(c), we also fit the decay of the computational-subspace population $P_{\mathcal{X}_1}$ using $C_\mathrm{SRB/IRB} + D_\mathrm{SRB/IRB} \cdot l_\mathrm{SRB/IRB}^m$, from which we calculate the leakage rate
\begin{align}
    L_\mathrm{SRB/IRB} = \frac{1 - C_\mathrm{SRB/IRB}}{1 - l_\mathrm{SRB/IRB}}.
\end{align}
Using these results, the CZ-gate error and leakage are respectively given as
\begin{align}
    r_\mathrm{CZ} &= \frac{d-1}{d} \left(1 - \frac{\lambda_\mathrm{IRB}}{\lambda_\mathrm{SRB}} \right), \label{eq:r_cz} \\
    L_\mathrm{CZ} &= 1 - \frac{1 - L_\mathrm{IRB}}{1 - L_\mathrm{SRB}}.
\end{align}
Note that Eq.\,\eqref{eq:r_cz} underestimates the impact of leakage~\cite{Wood_Leak}. Therefore, we calculate the total gate error as $e_\mathrm{CZ} = r_\mathrm{CZ} + L_\mathrm{CZ} / d$. From the experimental data shown in Figs.\,\ref{fig:SRB_IRB_2Q}(b) and (c), we obtain a CZ-gate fidelity of $F_\mathrm{CZ} = 1 - e_\mathrm{CZ} = 99.7(1)\%$, which corresponds to the first data points in Figs.\,\ref{fig:SRB_IRB_2Q}(d) and (e) immediately after the calibration. Furthermore, to assess the stability of the calibrated gate, we monitor the CZ gate error and leakage rate over approximately one day, as shown in Figs.\,\ref{fig:SRB_IRB_2Q}(d) and (e). For several hours after calibration, both quantities almost achieve the coherence limit obtained from the error model in Appendix~\ref{sec: error_model} and the independently measured coherence times in Appendix~\ref{sec: coherence}. After around ten hours, however, the gate error and leakage rate occasionally degrade. The simultaneous degradation of both metrics suggests that fluctuations in nearby two-level systems~\cite{TLS2019, Carroll2022} temporarily decrease device coherence, or that drifts in the drive frequency or amplitude shift the system away from the calibrated optimal point due to temperature changes in the setup. A detailed investigation of the temporal stability of gate fidelity is left for future work.

\section{\label{sec:ped_demo}Partial erasure-error detection}
\begin{figure}
    \centering
    \includegraphics{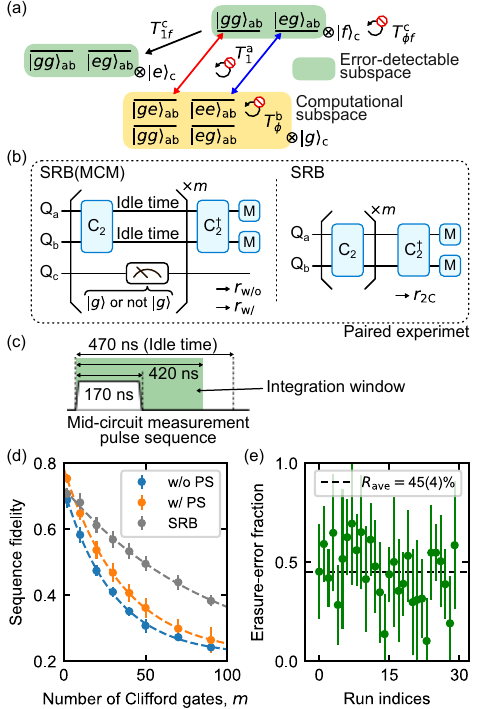}
    \caption{Partial erasure-error detection~(PED). (a)~Schematic illustration of decoherence-induced errors detectable in the TIP scheme. The computational subspace is highlighted in yellow, and the subspace involving coupler-transmon excitations (detectable erasure errors) is shaded in green. Decay processes marked with a no-go symbol induce dephasing and stop the $\fggetxt$ Rabi oscillation.
    (b)~SRB sequences with and without mid-circuit measurements (MCM). Exponential fits to SRB and SRB(MCM) results with and without post-selection using MCM outcomes yield $r_\mathrm{2C}$, $r_\mathrm{w/}$, and $r_\mathrm{w/o}$, respectively.
    (c)~MCM pulse consisting of a 170-ns flat-top microwave pulse with 10-ns Gaussian edges, followed by a 300-ns buffer before the next gate, giving an effective idle time of 470~ns for the data qubits. For each RB protocol, 10 random Clifford sequences are generated; each sequence is executed 1000 times, and results are averaged to obtain sequence fidelities.
    (d)~Sequence fidelities obtained in SRB and SRB(MCM). Blue: SRB(MCM) without post-selection. Orange: SRB(MCM) with post-selection on the coupler remaining in $\ket{g}$ throughout all MCMs. Gray: SRB without MCM pulses but with an added idle time. Dashed lines show the corresponding exponential-decay fits used to extract the two-qubit Clifford errors. The estimated values are $r_\mathrm{w/o}=2.44(9)\%$, $r_\mathrm{w/}=2.04(5)\%$, and $r_\mathrm{2C}=0.8(1)\%$.
    (e)~Erasure-error fraction estimated from SRB and SRB(MCM). Thirty independent measurements are performed over approximately 10~hours. Accounting for per-point uncertainties, the average erasure-error fraction is $R_\mathrm{ave}=45(4)\%$. In (d) and (e), error bars represent $\pm1\sigma$ standard errors.}
    \label{fig:erasure_fraction}
\end{figure}
We experimentally demonstrate that a subset of decoherence errors during the TIP gate is detected as erasure errors by measuring the coupler immediately after the gate is applied. As only a portion of the total decoherence is detectable in this scheme, we refer to this method as partial erasure-error detection (PED). The TIP gate populates the coupler's $\ket{f}_\mathrm{c}$ state. Thus, decoherence during the gate can leave residual population in the coupler's excited states. As shown in Fig.\,\ref{fig:erasure_fraction}(a), when we consider decoherence to first-order, analytical calculations (Appendix~\ref{sec: error_model}) identify four processes that produce coupler leakage: (i) energy relaxation from $\ket{e}_\mathrm{a}$ to $\ket{g}_\mathrm{a}$, which induces decoherence in the subspaces spanned by $\ket{ggf}$, $\ket{geg}$, $\ket{egf}$, and $\ket{eeg}$ during the CZ gate; (ii) energy relaxation from $\ket{f}_\mathrm{c}$ to $\ket{e}_\mathrm{c}$, which leaks into subspaces containing the first excited state of Q$_\mathrm{c}$; (iii) dephasing in the $\ket{g}_\mathrm{b}$–$\ket{e}_\mathrm{b}$ subspace; and (iv) dephasing in the $\ket{e}_\mathrm{c}$–$\ket{f}_\mathrm{c}$ subspace, both of which dephase the TIP Rabi oscillations shown as the red and blue arrow in Fig.\,\ref{fig:erasure_fraction}(a) and leave residual population in subspaces containing the second excited state of Q$_\mathrm{c}$. By detecting the resulting residual excitation of Q$_\mathrm{c}$ after the CZ gate, we identify gate failures. Since these error signals include location information, they can be treated as erasure errors.

\begin{figure}
    \centering
    \includegraphics{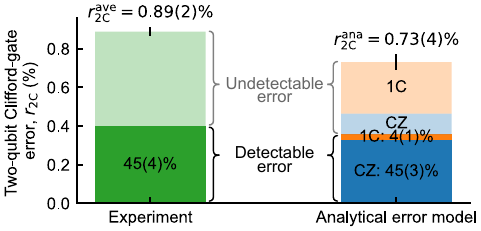}
    \caption{Breakdown of detectable and undetectable errors in the two-qubit Clifford gates, comparing the experimental results with the analytical estimate $r_\mathrm{2C}^\mathrm{ana} = 1.5r_\mathrm{CZ}^\mathrm{ana} + 5r_\mathrm{1C}^\mathrm{ana}$. Experimental data show that $45(4)\%$ of the total error is detectable with PED. The analytical model predicts that $49(3)\%$ of the two-qubit Clifford-gate error is PED-detectable, with contributions of $45(3)\%$ from the CZ-gate error and $4(1)\%$ from the single-qubit Clifford-gate error. Error bars represent $\pm1\sigma$ standard errors.}
    \label{fig:fraction_budgeting}
\end{figure}

In the context of QEC with erasure errors, it is essential to estimate the erasure-error fraction, i.e., the fraction of total errors that are erasures~\cite{Wu2022, KubicaErasure}. To this end, we use SRB together with a modified sequence shown in Fig.\,\ref{fig:erasure_fraction}(b): a dispersive readout of the coupler transmon is inserted after each two-qubit Clifford gate as a mid-circuit measurement (MCM). We refer to the modified SRB as SRB(MCM). As shown in Fig.\,\ref{fig:erasure_fraction}(c), the MCM consists of a 170-ns flat-top Gaussian pulse including 10-ns rising and falling Gaussian edges, followed by a 300-ns buffer to relax the readout resonator. During this 470-ns period, no operations are performed on Q$_\mathrm{a}$ or Q$_\mathrm{b}$. We perform 30~paired experiments—SRB without MCM and SRB(MCM) with MCM—and representative results are shown in Fig.\,\ref{fig:erasure_fraction}(d). In SRB(MCM), the two-qubit Clifford gate error $r_\mathrm{w/}$ is obtained by fitting the sequence fidelity post-selected on the runs where all MCM outcomes indicate the coupler in $\ket{g}_\mathrm{c}$. In contrast, $r_\mathrm{w/o}$ is obtained by fitting the fidelity calculated from all shots, without post-selection. For clarity, we denote the Clifford gate errors obtained from single- and two-qubit SRB, calculated using Eq.\,\eqref{eq:r_srb}, as $r_\mathrm{1C}$ and $r_\mathrm{2C}$, respectively. In addition, the single-qubit Clifford gate error used in subsequent calculations is the experimental value obtained from simultaneous RB in Appendix~\ref{sec: 1QRB}. Although the PED can mislabel events due to readout error, the probability of these mislabelings is small and of the order of the assignment readout infidelity (i.e., preparing the coupler in $\ket{e}$ or $\ket{f}$ but assigning it as $\ket{g}$). Therefore, the first-order expression for the erasure error fraction regarding small quantities (the probability of mislabeling detectable errors as no error and the gate errors) is as follows: 
\begin{align}
    R_\mathrm{1st} = \frac{\Delta r_\mathrm{PED} - p'_\mathrm{idle}}{r_\mathrm{2C}}, \label{eq:frac}
\end{align}
where $\Delta r_\mathrm{PED} = r_\mathrm{w/o} - r_\mathrm{w/}$. We assume independent single- and two-qubit error events. The quantity $p'_\mathrm{idle}$ in the numerator of Eq.\,\eqref{eq:frac} represents the detectable error during the idle time associated with the MCM pulse. Since this error is not included in the two-qubit Clifford gate error, the numerator indicates that this effect must be removed when estimating the erasure error fraction for the two-qubit gate error. The dominant detectable error in MCM is expected to be a phase error induced by the ZZ interactions between the coupler and data transmons when the coupler is excited. This error is estimated to be 0.056(8)\% from an analytical model using the coupler’s energy-relaxation time and steady-state thermal excitation probability (see Appendix~\ref{sec: fraction_cal} for details). Using Eq.\,\eqref{eq:frac} and the estimated detectable idling error, the erasure-error fraction is extracted from the data as shown in Fig.\,\ref{fig:erasure_fraction}(e). The error bars of individual data points are relatively large because the standard error of the two-qubit Clifford gate error in Eq.\eqref{eq:frac} is comparable to the detectable error. As a result, error propagation produces a large uncertainty in the calculated error fraction. To reduce this effect, we average 30\,paired-experimental runs, yielding an erasure fraction of $R_\mathrm{ave}=45(4)\%$. The result shows that roughly half of the errors are experimentally detectable, as illustrated by the green stacking bars in Fig.\,\ref{fig:fraction_budgeting}.

Finally, we compare the obtained erasure-error fraction with the analytical prediction based on the independently measured coherence times and analytical error models. The two-qubit Clifford gate error is decomposed as $r_\mathrm{2C}=1.5r_\mathrm{CZ} + 5r_\mathrm{1C}$, from which the analytical fraction of detectable two-qubit Clifford gate errors is given by
\begin{align}
    R' = \frac{1.5r_\mathrm{CZ}^\mathrm{ana}R_\mathrm{CZ} + 5r_\mathrm{1C}^\mathrm{ana}R_\mathrm{1C}}{1.5r_\mathrm{CZ}^\mathrm{ana} + 5r_\mathrm{1C}^\mathrm{ana}}.\label{eq:frac_ana}
\end{align}
Here, $R_\mathrm{CZ}$ and $R_\mathrm{1C}$ denote the PED-detectable fractions for the CZ and single-qubit Clifford gates. In addition, $r_\mathrm{CZ}^\mathrm{ana}$ and $r_\mathrm{1C}^\mathrm{ana}$ are the analytically calculated errors for the respective gates by assuming the performances of the gates are limited by qubit coherence (see Appendix~\ref{sec: error_model} for more details). Using the experimentally measured coherence times, Eqs.\,\eqref{eq:fraction_1Q} and\,\eqref{eq:tip_cz_fraction_ana}, we estimate $R_\mathrm{CZ}=75(4)\%$ and $R_\mathrm{1C}=10(1)\%$. Substituting these values into Eq.\,\eqref{eq:frac_ana} together with the analytical values of $r_\mathrm{CZ}^\mathrm{ana}$ and $r_\mathrm{1C}^\mathrm{ana}$, we estimated the detectable fraction of two-qubit Clifford errors to be $R' = 49(3)\%$. The error budget is summarized in Fig.\,\ref{fig:fraction_budgeting}. The analytical erasure-error fraction agrees with the experimental estimate of $45(4)\%$ within uncertainties, supporting the validity of the analytical model. Note that the finite detectable single-qubit gate errors arise from phase noise induced by excitation of the coupler and ZZ interactions between Q$_\mathrm{c}$ and the data transmons. Because these errors involve coupler excitation, they are also detectable by the PED.

With the current device, roughly half of the two-qubit Clifford gate errors are detectable. Increasing this fraction is an important future target. The analytic error budget in Fig.\,\ref{fig:fraction_budgeting} shows that single-qubit Clifford gate errors account for a large portion of the undetectable errors, resulting from a relatively long 100-ns single-qubit Clifford gate in our experiments. We expect that the detectable fraction of error can be increased by shortening the duration of the single-qubit gate and improving the coherence time of the data transmons. In addition, fast and high-fidelity readout via improved readout circuits~\cite{PeterPRXQ} would also help improve detection efficiency and reduce spurious idle errors.

\section{\label{sec:conc_disc}CONCLUSION and Discussion}
In this work, we proposed and experimentally demonstrated an all-microwave CZ gate that exploits auxiliary levels of a fixed-frequency transmon coupler and the $\fggetxt$ transition between one of the data transmons and the coupler. Because the transition frequency depends on the state of the other data transmon, we harnessed constructive interference of geometric phases to realize the CZ gate without requiring full-contrast Rabi oscillations. Notably, although the detuning between the data transmons in our device is about 500 MHz, far outside the straddling regime, the CZ-gate time is 140~ns, placing it among the fastest reported all-microwave two-qubit gates. The required drive strength, $\Omega_\mathrm{d}/2\pi \sim 230$~MHz, is relatively large; we attribute this mainly to fabrication-induced deviations and smaller interaction strengths relative to the target design value. With further optimization of the fabrication and design parameters, we expect that a drive strength of $\Omega_\mathrm{d}/2\pi \sim 100$ MHz will enable a CZ gate with a gate time of approximately 100~ns. A detailed discussion of design parameters can be found in Appendix~\ref{sec: Circ}.

Another notable feature of the Transmon-Induced Phase~(TIP) gate is the simplicity of the pulse shape required to implement the CZ gate. In the cross-resonance~(CR) and resonator-induced phase gates, the complexity of the underlying mechanisms often necessitates dynamical decoupling or composite pulses to approach the coherence limit. In contrast, here we achieve the highest CZ-gate fidelity of $99.7(1)\%$, close to the coherence limit, using only the simple flat-top raised-cosine pulse. This simplicity stems from the fact that the CZ operation involves only two geometric paths on the Bloch sphere conditioned on the control-qubit states. Analytical error modeling indicates that the fidelity of our TIP gate is primarily limited by the coherence time of the coupler transmon. Therefore, improving the coupler coherence will provide even higher fidelities.

We also proposed and experimentally verified the partial erasure-error detection (PED) protocol, in which the coupler transmon is measured after each two-qubit Clifford gate to identify a fraction of gate failures as erasure errors. The experimentally obtained erasure fraction of $45(4)\%$ agrees well with the analytically estimated $49(3)\%$ based on the relaxation and dephasing error model, confirming the validity of our description of the underlying error processes. Because the PED only requires the nonlinear coupler that can be measured, it can be implemented across various architectures, including those employing tunable-frequency couplers. The ability to flag a substantial portion of gate errors as erasures thus provides a direct hardware-level interface to erasure-aware quantum-error-correction protocols, opening a promising route toward further suppression of logical error rates in scalable superconducting quantum processors.

A key challenge for scalable devices based on the TIP scheme with fixed-frequency transmons is to develop frequency-allocation strategies~\cite{Hertzberg2021, PhysRevResearch.4.023079, PhysRevApplied.21.024035, Zhang_allocation} that minimizes frequency collisions. Because the TIP gate operates over a broad range outside the straddling regime, it is expected to yield higher fabrication tolerance than approaches based on the CR interaction. Although the present TIP gate employs a simple flat-top pulse, the waveform can be further refined through techniques such as DRAG~\cite{DRAG, Li2024} or detuning-robust shaping~\cite{PhysRevA.109.012616} with relatively minor additional complexity. Combining these control refinements with the demonstrated PED should enable all-microwave control with both higher fidelity and improved robustness to temporal frequency fluctuations, advancing wiring-efficient, scalable superconducting quantum computers.

\begin{acknowledgments}
The authors acknowledge the Superconducting Quantum Electronics Research Team, the Superconducting Quantum Computing System Research Unit, and the Semiconductor Science Research Support Team at RIKEN for their support in device fabrication at the RIKEN Nanoscience Joint Laboratory, H. Terai and Y. Hishida for providing the TiN films, K. Kodama, T. Sugiyama, H. Goto and Y. Tabuchi for fruitful discussions, and K. Kikuchi for supporting the measurement system setup. This research has been supported by funding from JST PRESTO (Grant Number JPMJPR25F3), JST ERATO (Grant Number JPMJER2302), JST CREST (Grant Number JPMJCR24I5), JSPS KAKENHI (Grant Numbers JP24H00832 and JP24K23038), and MEXT Q-LEAP (Grant Number JPMXS0118068682).
\end{acknowledgments}

\section*{Data availability}
The datasets used and/or analyzed during the current study are available from the corresponding author on reasonable request.

\appendix

\section{Fabrication}
The fabricated superconducting circuits are based on a TiN superconducting film, which offers high-performance microwave circuits~\cite{Tominaga2025}. In the fabrication process, a TiN film of around 100-nm thick is initially sputtered on a precleaned high-resistivity ($>$20~k$\Omega \cdot$cm) (100)-oriented silicon wafer at 850${}^\circ$C. Then, the resonators, transmon capacitors, and control lines are patterned through \mbox{photolithography}. After the development of photoresist, the exposed TiN film is subjected to reactive ion etching employing CF$_4$ gas. Following the wafer-cleaning procedure involving organic remover, oxygen plasma, and hydrofluoric acid, we make Manhattan-type Josephson junctions with the \textit{in-situ} bandage technique~\cite{insitu_bandage} through aluminum deposition and lift-off, employing electron-beam lithography. Finally, the wafer is diced into \mbox{$5$$\times$$5$-mm$^2$} chips, which are wire-bonded to a home-designed printed circuit board.

\section{Measurement setup}
\begin{figure}
    \centering
    \includegraphics{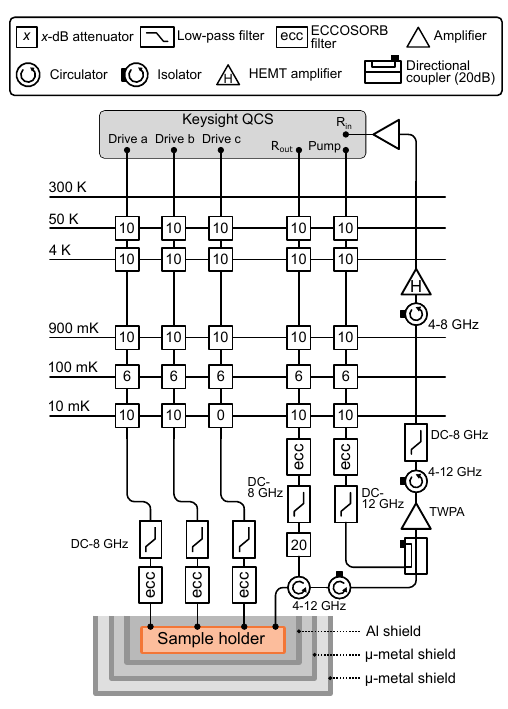}
    \caption{Schematic of the wiring in the dilution refrigerator and the room‑temperature measurement setup. Attenuators at each temperature stage suppress thermal noise from higher-temperature stages. At the 10-mK stage, infrared-absorbing \mbox{ECCOSORB} filters and 8-GHz low-pass filters are mounted above the three-layer magnetic shield. A traveling-wave parametric amplifier~(TWPA) followed by a high-electron-mobility-transistor~(HEMT) amplifier amplifies the readout signal before transmitting it to room temperature. The gold‑plated copper sample holder is mounted inside the three‑layer magnetic shield.}
    \label{fig:wiring}
\end{figure}

As shown in Fig.\,\ref{fig:wiring}, each input line of the dilution refrigerator has about 54-dB attenuation at 8~GHz, including the cable loss. Each drive line also has an \mbox{ECCOSORB} filter, an 8-GHz low-pass filter, and an extra 6-dB~(20\nobreakdash-dB) attenuator for the qubit\,(resonator) drive line. Note that, for the coupler drive line, a 10-dB attenuator at the \mbox{10-mK} stage is replaced with a 0-dB attenuator. The sample is mounted in a light-tight package placed inside a three-layer magnetic shield and cooled down to $\sim$10\,mK. Microwave pulses are generated by the arbitrary waveform generator (Keysight M5300). The reflection pulses from the readout resonators are amplified by a traveling-wave parametric amplifier (TWPA) at the 10-mK stage and then by a low-noise HEMT amplifier at the 4-K stage. The amplified readout signals are demodulated to IQ points for data processing by the down\nobreakdash-converter and digitizer (Keysight M5201 and M5200).

\section{Readout calibration}
\begin{figure}
    \centering
    \includegraphics{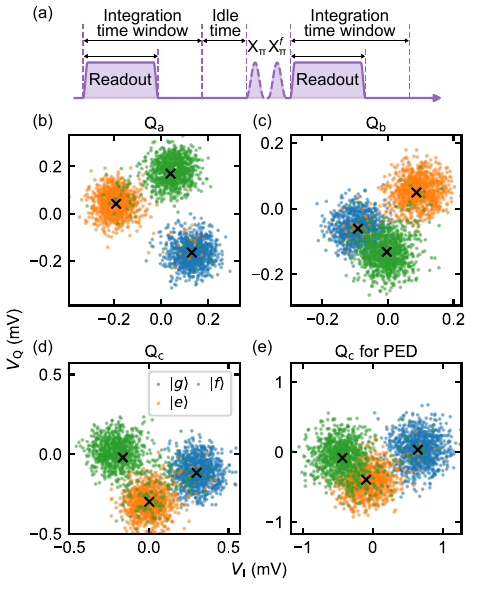}
    \caption{(a) Pulse sequence used to evaluate readout fidelity. The first pulse verifies, through post-selection, that the transmon is in the ground state, followed by a state-preparation pulse and a second readout. The integration time is measured from the beginning of the readout pulse, and a 200-ns idle time is inserted between pulses.
    (b,c,d) IQ distributions of the readout signals for the three states of Q$_\mathrm{a}$, Q$_\mathrm{b}$, and Q$_\mathrm{c}$, respectively. The black cross marks denote the reference points of each state. The measured signal is assigned to a state label according to the nearest reference point.
    (e) Example of the readout result of Q$_\mathrm{c}$ using the calibrated readout pulse for PED measurements.}
    \label{fig:readout_condition}
\end{figure}

\begin{table}[t]
    \caption{Transmon assignment-probability matrices. The last row shows the assignment matrix obtained under the readout-pulse condition for the PED measurements. Values in parentheses denote the $\pm1\sigma$ standard error.}
    \centering
    \begin{ruledtabular}
    \begin{tabular}{lcccc}
    & &\multicolumn{3}{c}{Readout state} \\
                           & & $\ket{g}$ & $\ket{e}$ & $\ket{f}$ \\ \hline
    
    Q$_\mathrm{a}$, $F_\mathrm{assign}=0.975(3)$ \\ 
    Target state & $\ket{g}$ & $0.9995(4)$ & $0.0003(2)$ & $0.0002(2)$ \\ 
    & $\ket{e}$ & $0.023(5)$ & $0.968(7)$ & $0.009(5)$ \\ 
    & $\ket{f}$ & $0.023(1)$ & $0.031(7)$ & $0.958(7)$ \\ \hline

    Q$_\mathrm{b}$, $F_\mathrm{assign}=0.909(7)$ \\ 
    Target state & $\ket{g}$ & $0.909(9)$ & $0.0074(5)$ & $0.084(10)$ \\ 
    & $\ket{e}$ & $0.042(10)$ & $0.941(20)$ & $0.018(10)$ \\ 
    & $\ket{f}$ & $0.093(3)$ & $0.028(2)$ & $0.879(3)$ \\  \hline

    Q$_\mathrm{c}$, $F_\mathrm{assign}=0.916(9)$ \\ 
    Target state & $\ket{g}$ & $0.979(5)$ & $0.021(4)$ & $0.0008(4)$ \\ 
    & $\ket{e}$ & $0.057(20)$ & $0.918(24)$ & $0.024(4)$ \\ 
    & $\ket{f}$ & $0.049(5)$ & $0.102(12)$ & $0.850(14)$ \\  \hline

    Q$_\mathrm{c}$, PED \\ 
    Target state & $\ket{g}$ & $0.979(4)$ & $0.019(4)$ & $0.0014(3)$ \\ 
    & $\ket{e}$ & $0.026(6)$ & $0.848(8)$ & $0.126(3)$ \\ 
    & $\ket{f}$ & $0.017(1)$ & $0.142(6)$ & $0.841(7)$ \\ 
    \end{tabular}
    \end{ruledtabular}
    \label{tab:assignment_mat}
\end{table}

\begin{table}[t]
    \caption{Readout parameters: For the readout pulse duration and integration-time window of the $\Qc$, values in parentheses indicate those used for PED. Values in other parentheses denote the $\pm1\sigma$ standard error.}
    \centering
    \begin{ruledtabular}
    \begin{tabular}{l@{\hspace{0.4em}}c@{\hspace{0.4em}}c@{\hspace{0.4em}}c}
                             & Q$_\mathrm{a}$ & Q$_\mathrm{b}$ & Q$_\mathrm{c}$ \\ \hline
    Readout-pulse frequency (GHz) & $7.734$ & $7.881$ & $7.932$ \\
    Dispersive shift, $\chi/2\pi$ (MHz) & $1.11$ & $0.95$ & $0.82$ \\
    External coupling, $\kappa_\mathrm{r}/2\pi$ (MHz) & $2.23$ & $0.94$ & $1.95$ \\
    Readout pulse duration (ns) & $1620$ & $1620$ & $620\,(170)$ \\
    Integration-time window  (ns) & $2200$ & $2200$ & $1000\,(420)$ \\
    Thermal $\ket{e}+\ket{f}$ population & $0.03(3)$ & $0.16(3)$ & $0.13(1)$ \\
    \end{tabular}
    \end{ruledtabular}
    \label{tab:ro_fid}
\end{table}
We set the readout‑pulse conditions by sweeping the pulse frequency and amplitude and selecting the operating point that maximizes the assignment fidelity. We measure the assignment fidelity using the pulse sequence in Fig.\,\ref{fig:readout_condition}(a). The first readout pulse verifies that the initial state is $\ket{g}$ by post-selection, after which we prepare one of the targets states, $\ket{g}$, $\ket{e}$, or $\ket{f}$, using combinations of X$_{\pi}$ and X$_{\pi}^f$ pulses. For ground‑state preparation, the preparation‑pulse amplitudes are set to zero. Figures~\ref{fig:readout_condition}(b)–(e) show the post‑selected IQ distributions of the readout signals, including those used in the PED measurements, for each transmon conditioned on the prepared target states.

For each target state, we perform the pulse sequence 2000 times and calculate a probability $P_x(y)$, which denotes the probability of obtaining outcome $\ket{y}$ when the sequence for a target state $\ket{x}$ is executed. We repeat this procedure five times and obtain the assignment matrices and the assignment fidelities $F_\mathrm{assign}$, summarized in Table~\ref{tab:assignment_mat}. We calculate the assignment fidelity as $F_\mathrm{assign} = (P_g(g) + P_e(e) + P_f(f))/3$. The parameters of readout resonators and pulses and the steady-state thermal‑excitation probabilities of the $\ket{e}$ and $\ket{f}$ states for each transmon are summarized in Table~\ref{tab:ro_fid}.

\section{\label{sec: coherence}Coherence-time measurement}
\begin{figure}
    \centering
    \includegraphics{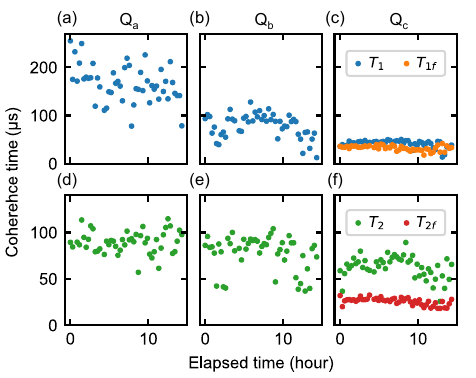}
    \caption{Long-time stability of the qubit coherence times. (a,b,c)~Energy-relaxation times $T_1$ of Q$_\mathrm{a}$, Q$_\mathrm{b}$, and Q$_\mathrm{c}$, respectively. The relaxation times from $\ket{e}$ to $\ket{g}$ are shown in blue. For Q$_\mathrm{c}$, the relaxation time $T_{1f}$ from $\ket{f}$ to $\ket{e}$ is also shown in orange dots.
    (d,e,f) Dephasing times. The dephasing times $T_2$ shown in green dots are obtained from the standard Hahn-echo sequence performed in the $\ket{g}$–$\ket{e}$ subspace. The dephasing time $T_{2f}$ in the $\ket{e}$–$\ket{f}$ subspace is also shown in red dots.}
    \label{fig:coherence_paper}
\end{figure}
\begin{figure}
    \centering
    \includegraphics{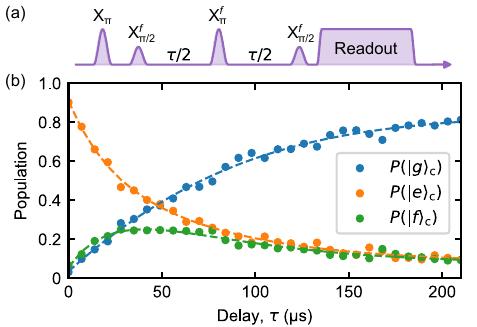}
    \caption{Hahn-echo measurement performed in the $\ket{e}$–$\ket{f}$ subspace of Q$_\mathrm{c}$. (a)~Pulse sequence. (b)~Typical experimental results. The blue, orange, and green dots represent the populations of $\ket{g}$, $\ket{e}$, and $\ket{f}$, respectively. The dashed lines indicate fits based on the master-equation model.}
    \label{fig:echo_q2_fit}
\end{figure}

Figure~\ref{fig:coherence_paper} summarizes measured energy‑relaxation and dephasing times for the device. We estimate the energy‑relaxation time $T_1$ for the $|e\rangle$$\rightarrow$$|g\rangle$ transition by preparing the transmon in $\ket{e}$ and fitting the exponential decay of the $\ket{e}$ population over time. We obtain the $\ket{f}$$\rightarrow$$\ket{e}$ relaxation time of Q$_\mathrm{c}$, $T_1^f$, by preparing $\ket{f}$ and fitting the decay of the $\ket{f}$ population. In this analysis, we neglect direct two‑photon decay from $\ket{f}$ to $\ket{g}$.

For dephasing in the $\ket{g}$–$\ket{e}$ subspace, we use a standard Hahn‑echo sequence and fit the decay of the $\ket{g}$ population to an exponential curve to estimate the dephasing time. In the $\ket{e}$–$\ket{f}$ subspace, population leakage cannot be neglected. Thus, we fit the data by numerically simulating the pulse sequence shown in the inset of Fig.\,\ref{fig:echo_q2_fit}. For X$_\pi$, X$_\pi^f$, and X$_{\pi/2}^f$, we assume ideal unitary gates, and model the idle intervals before and after the $X_\pi^f$ gate by solving the master equation
\begin{align}
    \dot{\hat{\rho}}(t) = \sum_l\mathcal{L}_l\hat{\rho}(t), \label{eq:ME}
\end{align}
with $\mathcal{L}_l = \hat{L}_l \hat{\rho} \hat{L}_l^\dagger - \{\hat{L}_l^\dagger \hat{L}_l, \hat{\rho}\}/2$, where $\hat{L}_l$ is a jump operator for a decoherence process. The decoherence processes included are energy relaxations from $\ket{e}$ and from $\ket{f}$ and pure dephasing within the $\ket{e}$–$\ket{f}$ subspace, modeled as $\hat{L}_1 = \sqrt{\Gamma_1}\dyad{g}{e}$, $\hat{L}_{1f} = \sqrt{\Gamma_{1f}}\dyad{e}{f}$, and $\hat{L}_{\phi f} = \sqrt{2\Gamma_{\phi f}}\dyad{f}{f}$, respectively. From the fit, we calculate the effective dephasing time in the $\ket{e}$–$\ket{f}$ subspace as $T_{2f} = 1/(\Gamma_{\phi f} + \Gamma_{1f}/2)$. Using $\hat{\rho}(t)$ calculated from Eq.\,\eqref{eq:ME}, the population of each state is calculated as $P_m=\mathrm{Tr}[\hat{\rho} \hat{S}_m]~(m \in \{g, e, f\})$. Here, $\hat{S}_m$ is a POVM operator that accounts for readout errors and thermal excitations modeled as 
\begin{align}
    \hat{S}_m = \sum_{n\in\{g,e,f\}}M_{mn}\dyad{n}{n},
\end{align}
where $M_{mn}$ denotes the probability of reporting outcome $\ket{m}$ when the true state is $\ket{n}$. These satisfy the POVM condition of $\hat{S}_g + \hat{S}_e + \hat{S}_f = \hat{I}$.

Figure~\ref{fig:echo_q2_fit} shows a typical fit using this model. The fit parameters include $T_1$, $T_{\phi f}$, and the six coefficients $\{M_{mn}\}$ considering the POVM condition. For $T_{1f}$, we impose the constraint $T_{1f} = c_0 T_1$ with $c_0 = T_{1f}^0 / T_1^0$, where $T_{1f}^0$ and $T_1^0$ are the relaxation times fitted to the data taken just before the dephasing measurement.

\section{\label{sec: 1QRB}Single-qubit gate fidelity}
\begin{figure*}
    \centering
    \includegraphics{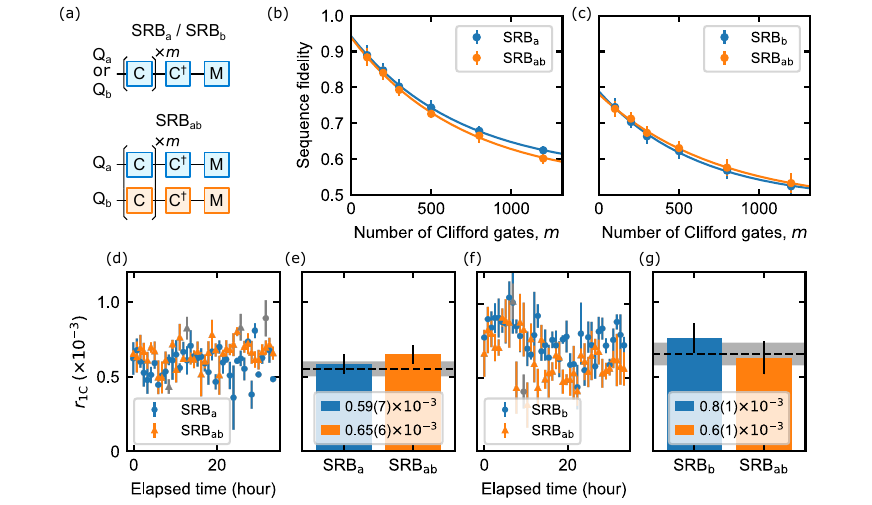}
    \caption{(a)~RB sequences with single‑qubit Clifford gates. We use standard randomized benchmarking (SRB) in two settings: individual SRB$_\mathrm{a}$ on Q$_\mathrm{a}$ and SRB$_\mathrm{b}$ on Q$_\mathrm{b}$, and simultaneous SRB$_\mathrm{ab}$ with both transmons. The sequence fidelity is calculated separately for Q$_\mathrm{a}$ and Q$_\mathrm{b}$. We average over 20 random RB sequences, and each sequence is repeated 1000 times.
    (b,c)~Representative individual SRB~(blue) and simultaneous SRB~(orange) results for Q$_\mathrm{a}$ and Q$_\mathrm{b}$. The error bars indicate $\pm1\sigma$ standard errors of the averaged measurement data.
    (d,f)~Single‑qubit Clifford‑gate errors~$r_\mathrm{1C}$ for Q$_\mathrm{a}$ and Q$_\mathrm{b}$, measured over approximately one day. Note that data points outside the inter-quartile range are considered outliers, as shown by the gray‑shaded points, and excluded from the analysis.
    (e,g)~Averages over the one‑day data for single‑qubit Clifford‑gate errors. Black solid error bars indicate $\pm1\sigma$ standard errors. Horizontal dashed lines and gray shaded bands indicate coherence limits from the analytical error model Eq.\,\eqref{eq:1Q_error_model} and their $\pm1\sigma$ standard error.}
    \label{fig:SRB_1Q}
\end{figure*}
Figure~\ref{fig:SRB_1Q} summarizes single‑qubit Clifford‑gate performance. To assess the impact of the residual ZZ interaction between Q$_\mathrm{a}$ and Q$_\mathrm{b}$, we perform the single-qubit individual and simultaneous standard RB~(SRB) on the two transmons. For SRB, the sequence fidelity is the population of $\ket{g}$. The single‑qubit Clifford‑gate error is calculated as $r_\mathrm{1C} = \frac{d-1}{d}(1-\lambda_\mathrm{SRB})$, using the fitting results. The long‑term measurements agree with the coherence limits predicted by the analytical error model Eq.\,\eqref{eq:1Q_error_model} within uncertainties.

\section{Leakage and survival probability in the PED experiment \label{sec: leak_suv_ped}}
\begin{figure}
    \centering
    \includegraphics{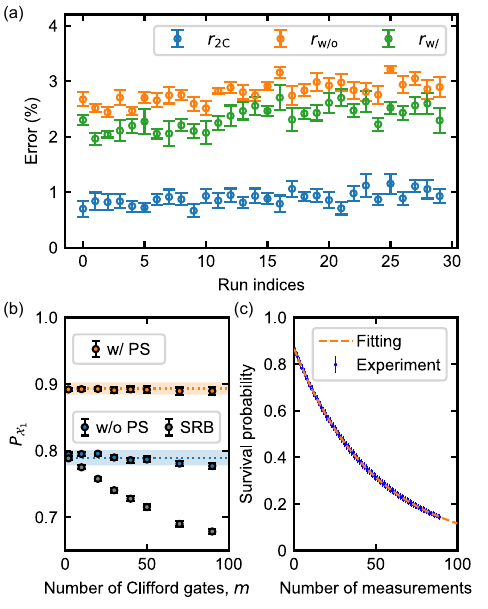}
    \caption{Dataset used to estimate the erasure-error fraction experimentally. (a)~Error rates obtained by fitting SRB and SRB(MCM) results with exponential decay models. (b)~Mean computational-subspace population over 30 runs for SRB and SRB(MCM). Orange, blue, and gray points denote SRB(MCM) and SRB results with (w/ PS) and without post-selection (w/o PS) using the PED outcomes, and without the MCM pulse (SRB), respectively. Horizontal dotted lines and shaded bands indicate the $m=0$ computational-subspace population estimated from the transmon thermal excitations and its $\pm 1\sigma$ standard error. With the PED, the first measurement removes initial thermal excitation of the coupler transmon, thereby increasing the computational-subspace population. (c)~Survival probability (acceptance probability) of single-shot trials after the post-selection. Blue points show the average over 30 runs with error bars. The orange dashed line shows the exponential fit $A(1-p_e)^m+B$, where $A$ and $B$ are fitting constants and $p_e$ denotes the probability that an erasure error is flagged. From the fit, these parameters are estimated to be $A=86.07(5)\%$, $B=0.53(6)\%$, and $p_e=2.029(3)\%$, respectively. In (a), (b), and (c), error bars represent $\pm1\sigma$ standard errors.}
    \label{fig: PED_leak_Ppos}
\end{figure}
Figure~\ref{fig: PED_leak_Ppos}(a) shows the dataset used to estimate the erasure-error fraction, obtained by exponential fits.
Figure~\ref{fig: PED_leak_Ppos}(b) presents the computational-subspace population $P_{\mathcal{\chi}_1}$ for SRB and SRB(MCM) with and without post-selection. The dotted lines indicate the $m=0$ populations estimated from each transmon’s thermal excitation and assignment fidelity. After post-selection, the population remains nearly unchanged from its initial value. Even without post-selection, the decay of the computational subspace population occurs more slowly compared to that of SRB without MCM pulses. This is likely due to readout-induced relaxation \cite{PhysRevLett.132.090602, PhysRevA.85.022305} or state transition \cite{MIST_disp} of the transmon, or due to both factors.
In Fig.~\ref{fig: PED_leak_Ppos}(c), the survival (acceptance) probability under post-selection is shown. An exponential fit yields a positive-outcome rate of $p_e = 2.029(3)\%$.

\section{\label{sec: Perturbation}Derivation of Eq.~(4)}
Here we explain the perturbative derivation of Eq.\,\eqref{eq:fgge_shift}. Comprehensive derivations of Eq.\,\eqref{eq:fgge} can be found in references~\cite{fgge_eq, Krinner_CZ}. As a starting point, we consider the Hamiltonian defined in Eq.\,\eqref{eq:H}. Since we have $g_\mathrm{ab} \ll g_\mathrm{ac}, g_\mathrm{bc}$, the transverse interaction between Q$_\mathrm{a}$ and Q$_\mathrm{b}$ can be neglected. Under this condition, the Hamiltonian in Eq.\,\eqref{eq:H} can be separated as
\begin{align}
    \hat{H}_0/\hbar &= \sum_i \left(\omega_i\Ap_i\Am_i + \frac{\alpha_i}{2}\Ap_i\Ap_i\Am_i\Am_i\right), \\
    \hat{V}/\hbar &= g_\mathrm{ac}(\Ap_\mathrm{a} \Am_\mathrm{c} + \Am_\mathrm{a}\Ap_\mathrm{c}) + g_\mathrm{bc}(\Ap_\mathrm{b} \Am_\mathrm{c} + \Am_\mathrm{b}\Ap_\mathrm{c}).
\end{align}
Following the procedure of the Schrieffer–Wolff transformation, the anti-Hermitian operator $\hat{A}$ is determined by solving the equation,
\begin{align}
    [\hat{H}_0, \hat{A}] + \hat{V} = 0. \label{eq:K}
\end{align}
Here, $\hat{A}$ is obtained by solving Eq.\,\eqref{eq:K} using a Python program~\cite{SS}. Defining $\hat{D}$ as the diagonal part of $[\hat{H}_\mathrm{V}, \hat{A}]/2$, we write the Hamiltonian diagonalized up to second-order perturbation as
\begin{align}
    \hat{H}' = \hat{H}_0 + \hat{D}. \label{eq:eff_H}
\end{align}
The frequency shift Eq.\,\eqref{eq:fgge_shift} can then be calculated as follows:
\begin{align}
    \Delta_\fgge &= \bra{ggf}\hat{H}'\ket{ggf}/\hbar - \bra{geg}\hat{H}'\ket{geg}/\hbar \notag\\
                 &- \bra{egf}\hat{H}'\ket{egf}/\hbar + \bra{eeg}\hat{H}'\ket{eeg}/\hbar.
\end{align}

\section{\label{sec: Circ}Device-design considerations}
\subsection{Electrodes layout}
\begin{figure}
    \centering
    \includegraphics{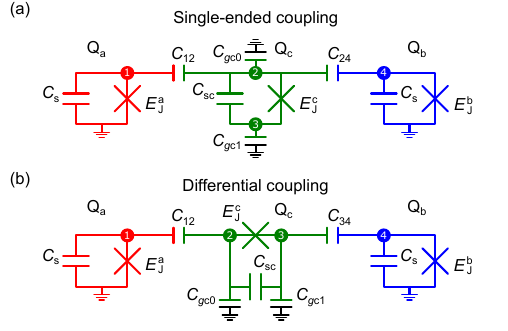}
    \caption{Simplified circuits of data transmons (red, blue) connected via a coupler (green). (a)~Single-ended coupling used in this work. (b)~Differential coupling, where the intrinsic static-ZZ suppression does not work; the data transmons couple to opposite electrodes of the coupler.}
    \label{fig: c_layout_not}
\end{figure}
Here, we explain how a circuit-layout choice suppresses residual ZZ interaction. For that, we use the simplified configurations shown in Figs.\,\ref{fig: c_layout_not}(a) and (b), where one electrode of each data transmon is grounded. In this work, we adopt the layout shown in Fig.\,\ref{fig: design}(a) of the main text~[its simplified version is Fig.\,\ref{fig: c_layout_not}(a)], in which two data transmons are capacitively coupled to the same electrode of the coupler transmon; we refer to these layouts as single-ended coupling. In this layout, the residual ZZ interaction between $\Qa$ and $\Qb$ can be suppressed when the coupler transmon has the highest transition frequency. On the other hand, for the circuit shown in Fig.\,\ref{fig: c_layout_not}(b), the data transmons are capacitively coupled to opposite electrodes of the coupler, and we call it differential coupling. The suppression of the residual ZZ interaction by reducing the net transverse interaction is ineffective in this layout and parameter regime because the transverse interaction $g_\mathrm{bc}$ is negative. 

To confirm the signs of the transverse interactions between transmons, we analyze each circuit layout and derive expressions for the transverse interactions. First, the capacitance matrix of the circuit in Fig.\,\ref{fig: c_layout_not}(a) is
\begin{align}
    \mathbf{C}_\mathrm{se} = 
    \left[\begin{matrix}
        C_\mathrm{11}^\mathrm{se} & - C_\mathrm{12} & 0 & 0\\
        - C_\mathrm{12} & C_\mathrm{22}^\mathrm{se} & - C_\mathrm{sc} & - C_\mathrm{24}\\
        0 & - C_\mathrm{sc} & C_\mathrm{33}^\mathrm{se} & 0\\
        0 & - C_\mathrm{24} & 0 & C_\mathrm{44}^\mathrm{se}
    \end{matrix}\right],
\end{align}
where
\begin{align}
    C_{11}^\mathrm{se} &= C_\mathrm{12} + C_\mathrm{s}, \\
    C_{22}^\mathrm{se} &= C_\mathrm{12} + C_\mathrm{24} + C_\mathrm{gc0} + C_\mathrm{sc}, \\
    C_{33}^\mathrm{se} &= C_\mathrm{gc1} + C_\mathrm{sc}, \\
    C_{44}^\mathrm{se} &= C_\mathrm{24} + C_\mathrm{s}.
\end{align}
For the circuit in Fig.\,\ref{fig: c_layout_not}(b), the capacitance matrix is
\begin{align}
    \mathbf{C}_\mathrm{diff} = 
    \left[\begin{matrix}
        C_\mathrm{11}^\mathrm{diff} & - C_\mathrm{12} & 0 & 0\\
        - C_\mathrm{12} & C_\mathrm{22}^\mathrm{diff} & - C_\mathrm{sc} & 0\\
        0 & - C_\mathrm{sc} & C_\mathrm{33}^\mathrm{diff} & - C_\mathrm{34}\\
        0 & 0 & - C_\mathrm{34} & C_\mathrm{44}^\mathrm{diff}
    \end{matrix}\right],
\end{align}
where
\begin{align}
    C_{11}^\mathrm{diff} &= C_\mathrm{12} + C_\mathrm{s}, \\
    C_{22}^\mathrm{diff} &= C_\mathrm{12} + C_\mathrm{gc0} + C_\mathrm{sc}, \\
    C_{33}^\mathrm{diff} &= C_\mathrm{34} + C_\mathrm{gc1} + C_\mathrm{sc}, \\
    C_{44}^\mathrm{diff} &= C_\mathrm{34} + C_\mathrm{s}.
\end{align}
For each node in the circuits, the node-flux variable is denoted by $\Phi_i$~($i \in \{\mathrm{1, 2, 3, 4}\}$), and introduce the coordinate transformation as $\mathrm{vec}(\Phi_\mathrm{a}, \Phi_\mathrm{c}, \Phi_\mathrm{cp}, \Phi_\mathrm{b}) = \mathbf{V}^{-1}\mathrm{vec}(\Phi_\mathrm{1}, \Phi_\mathrm{2}, \Phi_\mathrm{3}, \Phi_\mathrm{4})$, where
\begin{align}
    \mathbf{V} = 
    \left[\begin{matrix}
        1 & 0 & 0 & 0\\
        0 & 1 & 1 & 0\\
        0 & -1 & 1 & 0\\
        0 & 0 & 0 & 1
    \end{matrix}\right].
\end{align}
Introducing the normalized charge variable $n_j = \frac{1}{2e}\frac{\partial \mathcal{L}}{\partial \Phi_j}$~($j \in \{\mathrm{a, c, cp, b}\}$ and vector $\mathbf{n} = \mathrm{vec}(n_\mathrm{a}, n_\mathrm{c}, n_\mathrm{cp}, n_\mathrm{b})$, the respective charging energies $T_\mathrm{se/diff}$ of the circuits are written as \cite{RasmussenCircuitCompanion}
\begin{align}
    T_\mathrm{se/diff} = 4\frac{e^2}{2}\mathbf{n}^T\mathbf{K}_\mathrm{se/diff}^{-1}\mathbf{n}, \label{eq:charge_enes}
\end{align}
where $\mathbf{K}_\mathrm{se/diff} = \mathbf{V}^T\mathbf{C}_\mathrm{se/diff}\mathbf{V}$. Note that $n_\mathrm{cp}$ is a conservative quantity and does not affect the system dynamics. Thus, we ignore this variable hereafter~\cite{Chitta_2022}. The cross terms in Eq.\,\eqref{eq:charge_enes} correspond to the capacitive coupling energies between transmons mediated by the capacitors. For the single-ended coupling, the capacitive coupling energies for each transmon pair are written as
\begin{align}
    E_\mathrm{C,se}^\mathrm{ab} &= \frac{e^2C_{12}C_{24}(C_{g\mathrm{c1}}+C_{\mathrm{sc}})}{|\mathbf{C}_\mathrm{se}|}, \label{eq:EC_ab_se}\\
    E_\mathrm{C,se}^\mathrm{ac} &= \frac{e^2C_{12}C_{g\mathrm{c1}}(C_{24}+C_{\mathrm{s}})}{2|\mathbf{C}_\mathrm{se}|}, \\
    E_\mathrm{C,se}^\mathrm{bc} &= \frac{e^2C_{24}C_{g\mathrm{c1}}(C_{12}+C_{\mathrm{s}})}{2|\mathbf{C}_\mathrm{se}|}.
\end{align}
For the differential coupling, they become
\begin{align}
    E_\mathrm{C,diff}^\mathrm{ab} &= \frac{e^2C_{12}C_{34}C_{\mathrm{sc}}}{|\mathbf{C}_\mathrm{diff}|}, \\
    E_\mathrm{C,diff}^\mathrm{ac} &= \frac{e^2C_{12}(C_{34}C_{g\mathrm{c1}}+C_{34}C_{\mathrm{s}}+C_{g\mathrm{c1}}C_\mathrm{s})}{2|\mathbf{C}_\mathrm{diff}|}, \\
    E_\mathrm{C,diff}^\mathrm{bc} &= \frac{-e^2C_{34}(C_{12}C_{g\mathrm{c0}}+C_{12}C_{\mathrm{s}}+C_{g\mathrm{c0}}C_\mathrm{s})}{2|\mathbf{C}_\mathrm{diff}|}. \label{eq:EC_bc_diff}
\end{align}
In addition, the transverse interaction between transmons is given by
\begin{align}
    \hbar g_{ij} = E_\mathrm{C}^{ij}\left( \frac{E_\mathrm{J}^iE_\mathrm{J}^j}{4E_\mathrm{C}^iE_\mathrm{C}^j} \right)^{1/4}, \label{eq:coup_ana}
\end{align}
where $E_\mathrm{C}^i$ and $E_\mathrm{J}^i$ denote the charging and Josephson energies of $\mathrm{Q}_i$, respectively, and $E_\mathrm{C}^{ij}$ is the capacitive coupling energy between $\mathrm{Q}_i$ and $\mathrm{Q}_j$. Therefore, from Eqs.\,\eqref{eq:EC_ab_se}--\eqref{eq:EC_bc_diff} and Eq.\,\eqref{eq:coup_ana}, both the first and second terms in Eq.\,\eqref{eq:g_eff} are positive in the differential coupling layout with the parameter regime we consider. The direct and indirect transverse interactions do not cancel each other, and suppression of residual ZZ via Eq.\,\eqref{eq:zz} is inefficient. This fact motivates us to employ the single-ended coupling layout. Note that the signs of the derived interaction depend on the definitions of the positive direction of the branch voltages. However, the final physical results are invariant to those definitions. More systematic and detailed discussions treating the data transmons as the floating transmons can be found in Refs.~\citenum{SeteCoupler, Solgun}. It should be noted that, in Ref.~\citenum{SeteCoupler}, the signs of the transverse interaction differ from those used in this work due to the different definitions of voltage polarities between nodes, as mentioned above. This difference does not affect the final results.

\subsection{Parameter design}
\begin{figure}
    \centering
    \includegraphics{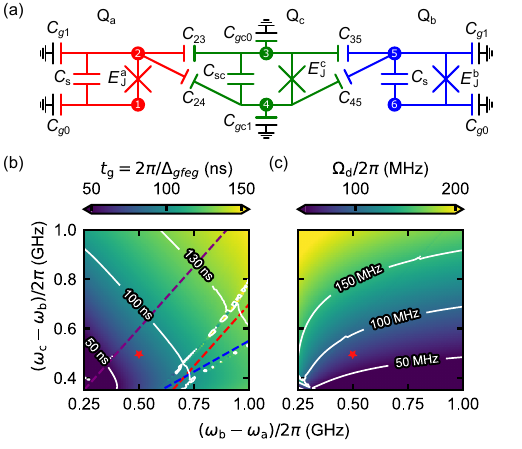}
    \caption{
    (a) Lumped-element circuit model for the numerical calculations. From the circuit geometry, we neglect the direct capacitance between the data transmons and the capacitances between the coupler transmon and nodes 1 and\,6.
    (b)~TIP-gate time obtained numerically. It is calculated as the inverse of the $\fggetxt$-transition frequency shift $\Delta_\fgge$ in Q$_\mathrm{b}$ and Q$_\mathrm{c}$. The experimental gate time is longer by the pulse-edge duration. The purple, red, and blue dashed lines indicate the resonance conditions between the $\fggetxt$ transition and the $\ket{eeg}$--$\ket{gfe}$, $\ket{egg}$--$\ket{ggh}$, and $\ket{eeg}$--$\ket{gff}$ transitions, respectively. To calculate the resonance conditions, the bare frequencies and anharmonicities of each transmon are used. (c)~Drive amplitude $\Omega_\mathrm{d}$ required to implement the TIP gate. Using the condition $\Omega_\fgge^g = \Omega_\fgge^e = \sqrt{3} \, \Delta_{gfeg}/2$ and Eq.\,\eqref{eq:fgge}, we calculate the required drive strength. In~(b) and (c), the horizontal axes are the detuning between the data transmons Q$_\mathrm{a}$ and Q$_\mathrm{b}$ with $\omega_\mathrm{a} / 2\pi = 4.5$~GHz fixed, and the vertical axes are the detuning between $\Qb$ and $\Qc$. The red star in each plot indicates the target parameters for the device fabrication in this study.}
    \label{fig: design_sweep}
\end{figure}

\begin{table}
    \caption{Parameter setting for the numerical studies. The anharmonicities $\alpha_i$~($i \in \{\mathrm{a},\mathrm{b},\mathrm{c}\}$) are not included directly in the optimization objective. They vary slightly depending on the optimized values of $C_{23}$ and $C_{35}(C_{45})$.}
    \begin{minipage}[t]{0.47\columnwidth}\vspace{0pt}
        \raggedright 
        \begin{tabular}{l@{\hspace{3em}}c}
        \hline\hline
        \multicolumn{2}{c}{Common} \\
        \hline
        $E_{J}^\mathrm{a}/\hbar$ (GHz) & Variable \\
        $E_{J}^\mathrm{b}/\hbar$ (GHz) & Variable \\
        $E_{J}^\mathrm{c}/\hbar$ (GHz) & Variable \\
        $-\alpha_\mathrm{a}/2\pi$ (MHz) & 201--206 \\
        $-\alpha_\mathrm{b}/2\pi$ (MHz) & 198--205 \\
        $-\alpha_\mathrm{c}/2\pi$ (MHz) & 289--314 \\
        $C_\mathrm{s}$ (fF) & 55 \\
        $C_\mathrm{g0}$ (fF) & 80 \\
        $C_\mathrm{g1}$ (fF) & 120 \\
        $C_\mathrm{sc}$ (fF) & 32 \\ \hline\hline
        \end{tabular}
    \end{minipage}%
    \hfill
    \begin{minipage}[t]{0.47\columnwidth}\vspace{0pt}
        \raggedleft 
        \begin{tabular}{l@{\hspace{3em}}c}
        \hline\hline
        \multicolumn{2}{c}{Single-ended} \\
        \hline
        $C_{23}$ (fF) & Variable \\
        $C_{24}$ (fF) & 0 \\
        $C_{35}$ (fF) & Variable \\
        $C_{45}$ (fF) & 0 \\
        $C_\mathrm{gc0}$ (fF) & 100 \\
        $C_\mathrm{gc1}$ (fF) & 60 \\ \hline\hline
        \\
        \hline\hline
        \multicolumn{2}{c}{Differential} \\
        \hline
        $C_{23}$ (fF) & Variable \\
        $C_{24}$ (fF) & 0 \\
        $C_{35}$ (fF) & 0 \\
        $C_{45}$ (fF) & Variable \\
        $C_\mathrm{gc0}$ (fF) & 75 \\
        $C_\mathrm{gc1}$ (fF) & 75 \\ \hline\hline
        \end{tabular}
    \end{minipage}
    \label{tab:params_sweep}
\end{table}
Next, we discuss how the gate speed and the required drive power depend on the design parameters in the single-ended layout and provide design guidelines for the TIP gate. For the circuit in Fig.\,\ref{fig: design_sweep}(a), circuit quantization yields the Hamiltonian,
\begin{align}
    \hat{H} = \sum_i \left(4E_\mathrm{C}^i\hat{n}_i^2 + E_\mathrm{J}^i\cos{\hat{\phi}_i}\right) + \sum_{i \neq j} 4E_\mathrm{C}^{ij}\hat{n}_i\hat{n}_j.
    \label{eq:h_circ}
\end{align}
Here, $\hat{n}_i$ and $\hat{\phi}_i$ are the normalized charge and flux operators satisfying $[\hat{\phi}_i, \hat{n}_j]=i\delta_{ij}$, respectively. From the coefficients of this Hamiltonian, the transmon frequencies and anharmonicities are approximated as follows~\cite{TmonAnalytical}:
\begin{align}
    \hbar\omega_i &= \sqrt{8E_\mathrm{C}^iE_\mathrm{J}^i} \notag \\
    &- E_\mathrm{C}^i\left( 1 + \frac{1}{2^2}\xi_i + \frac{21}{2^7}\xi_i^2 + \frac{19}{2^7}\xi_i^3 + \frac{5319}{2^{15}}\xi_i^4 \right), \label{eq:omega_ana}\\
    \hbar\alpha_i &= -E_\mathrm{C}^i\left( 1 + \frac{9}{2^4}\xi_i + \frac{81}{2^7}\xi_i^2 + \frac{3645}{2^{12}}\xi_i^3 + \frac{46899}{2^{15}}\xi_i^4 \right), \label{eq:alpha_ana}
\end{align}
where $\xi_i = \sqrt{2E_\mathrm{C}^i / E_\mathrm{J}^i}$. The transverse interaction between transmons is calculated by using Eq.\,\eqref{eq:coup_ana}.

In the numerical studies, we fix the lowest frequency at $\omega_\mathrm{a}/2\pi=4.5$ GHz, sweep $\Delta_\mathrm{ba}/2\pi=(\omega_\mathrm{b}-\omega_\mathrm{a})/2\pi$ over the range of 0.25--1.00~GHz, and sweep $\Delta_\mathrm{cb}/2\pi=(\omega_\mathrm{c}-\omega_\mathrm{b})/2\pi$ over the range of 0.35--1.00~GHz. For the couplings, we set the empirically chosen targets $|g_\mathrm{ac}/\Delta_\mathrm{ac}|=0.05$ and $|g_\mathrm{bc}/\Delta_\mathrm{bc}|=0.06$, which are expected to achieve gate times of 130~ns or less with drive amplitudes of 150~MHz or less. Note that, to prevent a larger drive amplitude, we need to increase the $\fggetxt$ pair interaction. On the other hand, if the state-dependent frequency shift $\Delta_\fgge$ becomes too large, achieving $\Omega_\fgge=\sqrt{3}\Delta_\fgge/2$ to implement the CZ gate also requires a larger drive amplitude. Therefore, we impose the condition that $|g_\mathrm{ac}/\Delta_\mathrm{ac}|<|g_\mathrm{bc}/\Delta_\mathrm{bc}|$. For the Single-ended layout, we numerically optimize the circuit variable set $\{ E_\mathrm{J}^\mathrm{a}, E_\mathrm{J}^\mathrm{b}, E_\mathrm{J}^\mathrm{c}, C_{23}, C_{35} \}$ so that the total mismatch from a target parameter set $\{ \omega_\mathrm{a}, \omega_\mathrm{b}, \omega_\mathrm{c}, g_\mathrm{ac}, g_\mathrm{bc} \}$ is less than 1 MHz. We use \mbox{scQubits}~\cite{Groszkowski2021scqubitspython} to compute the coefficients of the Hamiltonian in Eq.\,\eqref{eq:h_circ} numerically. The variables and constants used in the calculations are summarized in Table~\ref{tab:params_sweep}. We perform calculations for both the single-ended and differential coupling layouts, and parameters that differ between layouts are listed in separate tables for clarity. For the anharmonicities, we design the geometry to yield approximately $\alpha_\mathrm{c}/2\pi \sim -300$~MHz and $\alpha_\mathrm{a/b}/2\pi \sim -200$~MHz. The capacitance values between pads and to the ground plane are estimated with \mbox{COMSOL}~\cite{comsol}. Since the anharmonicities are not included directly in the optimization objective, they vary slightly with the optimized values of $C_{23}$ and $C_{35}$; the ranges are reported in Table~\ref{tab:params_sweep}. These ranges are similar for both layouts.

Using the optimized parameters, the resulting gate time and required drive power are estimated as shown in Figs.\,\ref{fig: design_sweep}(b) and (c). Here, the $\fggetxt$ Rabi-oscillation frequency $\Omega_\fgge^{g}$ and state-dependent frequency shift $\Delta_\fgge$ are numerically calculated. Across a wide parameter region with detunings approaching 1 GHz outside the straddling regime, the TIP gate is expected to operate with gate times of roughly 50--150~ns using drive strengths of 50--200~MHz. These plots indicate a practical parameter range for device design. For Fig.\,\ref{fig: design_sweep}(b), we observe regions where the white contour lines are distorted. These distortions arise from resonant crossings between the one-photon transition $\ket{eeg}$--$\ket{gfe}$ and the two-photon transitions $\ket{egg}$--$\ket{ggh}$ and $\ket{eeg}$--$\ket{gff}$ with the $\fggetxt$ transition. In the figure, the resonance conditions calculated from the bare frequencies and anharmonicities of each transmon are shown as dashed lines; these frequencies should be avoided in parameter design. A systematic study of how proximity to these resonances affects the TIP gate fidelity is an important direction for future work. Based on these observations, we design the device with $\omega_\mathrm{a}/2\pi=4.5$~GHz and $\Delta_\mathrm{ba}/2\pi=\Delta_\mathrm{cb}/2\pi=0.5$~GHz as indicated by the red stars in Figs.\,\ref{fig: design_sweep}(b) and\,(c).

\subsection{Intrinsic static-ZZ suppression}
\begin{figure}
    \centering
    \includegraphics{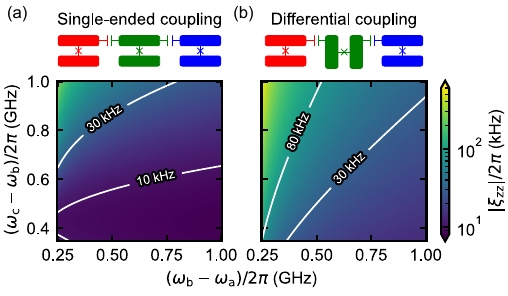}
    \caption{Residual ZZ interaction strength between $\Qa$ and $\Qb$ for (a) the single-ended coupling layout and (b) the differential coupling layout, computed from the numerical simulation results. The horizontal axes are the detuning between the data transmons with $\omega_\mathrm{a}/2\pi = 4.5$~GHz fixed, and the vertical axes are the detuning between Q$_\mathrm{b}$ and Q$_\mathrm{c}$.}
    \label{fig:zz_sim_se_diff}
\end{figure}
Next, using the parameter sets $\{ E_\mathrm{J}^\mathrm{a}, E_\mathrm{J}^\mathrm{b}, E_\mathrm{J}^\mathrm{c}, C_{23}, C_{35} \}$ obtained in the previous subsection, we evaluate the residual ZZ interactions between $\Qa$ and $\Qb$ for each layout. From these parameters, we calculate each transmon's frequency $\omega_i$ and anharmonicity $\alpha_i$, as well as the transverse interactions $g_\mathrm{ab}$, $g_\mathrm{ac}$, and $g_\mathrm{bc}$, and substitute them into the Hamiltonian~\eqref{eq:H}. We then numerically diagonalize the Hamiltonian and define the residual coupling as $\xi_\mathrm{ZZ} = \tilde{\omega}_{eeg} - \tilde{\omega}_{geg} - \tilde{\omega}_{egg}$~\cite{JAZZ1, fors2024comprehensiveexplanationzzcoupling}, where $\tilde{\omega}_{ijk}$ denotes an eigenfrequency of the dressed state closest to a bare state $\ket{ijk}$. The resulting residual ZZ interactions for the two layouts are shown in Figs.\,\ref{fig:zz_sim_se_diff}(a) and (b). The single-ended coupling layout suppresses the residual ZZ interaction by a factor of approximately 2 to 5 over a broad parameter region compared with the differential layout. This observation motivates our choice of the single-ended coupling layout in this work.

\section{\label{sec: error_model}Error modeling}
We derive error models for single-qubit gates and the TIP gate. We begin by outlining the method for deriving an incoherent error governed by a decoherence process $\mathcal{L}_v$, following Ref.~\citenum{FastLogic}. When the time-independent Hamiltonian implementing a target quantum gate is denoted by $\hat{H}_\mathrm{T}$, the master equation is given by
\begin{align}
    \dot{\hat{\rho}} = -\frac{i}{\hbar}[\hat{H}_\mathrm{T}, \hat{\rho}] + \mathcal{L}_v\hat{\rho}. \label{eq:ME_L}
\end{align}
We consider the case where the relaxation time $1/\Gamma_v$ associated with the decoherence process $\mathcal{L}_v$ is much longer than the duration $t_\mathrm{g}$ of the quantum gate, such that $\Gamma_v t_\mathrm{g} \ll 1$. In this regime, the density operator can be expanded as $\hat{\rho} = \hat{\rho}_0 + \hat{\rho}_1 + \cdots$. The zeroth-order time evolution is then given by
\begin{align}
    \dot{\hat{\rho}}_0 = -\frac{i}{\hbar}[\hat{H}_\mathrm{T}, \hat{\rho}_0]. \label{eq:ME_0th}
\end{align}
This corresponds to a unitary evolution. For the first-order terms, we obtain
\begin{align}
    \dot{\hat{\rho}}_1 = -\frac{i}{\hbar}[\hat{H}_\mathrm{T}, \hat{\rho}_1] + \mathcal{L}_v\hat{\rho}_0. \label{eq:ME_1st}
\end{align}
For a pair of eigenvectors $\ket{m}$ and $\ket{n}$ of $\hat{H}_\mathrm{T}$, the solution of Eq.\,(\ref{eq:ME_1st}) is expressed as
\begin{align}
    r_{mn}^{\psi,v} = \int_0^{t_\mathrm{g}} dt \bra{m}\mathcal{L}_v\hat{\rho}_0^{\psi}(t)\ket{n} e^{-i\omega_{mn}(t_\mathrm{g} - t)}, \label{eq:Sol_ME_1st}
\end{align}
where $r_{mn}^{\psi,v}$ is a matrix element of the first-order correction term of the density operator written as $\hat{\rho}_1^{\psi,v} = \sum_{mn} r_{mn}^{\psi,v} \dyad{m}{n}$, $\hbar \omega_m$ is the eigenvalue of $\hat{H}_\mathrm{T}$ corresponding to $\ket{m}$, $\omega_{mn} = \omega_m - \omega_n$, and $\hat{\rho}_0^{\psi}(t)$ is a solution of Eq.\,(\ref{eq:ME_0th}) for an initial state $\ket{\psi}$ in the computational subspace. Calculating Eq.\,(\ref{eq:Sol_ME_1st}) for all pairs of eigenvalues yields the first-order correction term of the density operator. Then, the first-order expression of the state fidelity is obtained as
\begin{align}
    F^{\psi}_v = \mathrm{Tr}[\hat{\rho}_0^\psi (\hat{\rho}_0^\psi + \hat{\rho}_1^{\psi,v})].
\end{align}
Gate errors due to given decoherence processes $\{\mathcal{L}_v\}$ are calculated by averaging over the set of input states in the computational subspace, $\ket{\psi} \in \{\ket{0}, \ket{1}, \ket{+}, \ket{-}, \ket{+i}, \ket{-i}\}^{\otimes q}$:
\begin{align}
    \bar{\epsilon}_v = \frac{1}{6^q} \sum_\psi \left(1 - F^{\psi}_v\right), \label{eq:inf_ana}
\end{align}
where $q=1$ and 2 for the single-qubit and two-qubit gates, respectively. By summing over all relevant decoherence processes, we obtain a total infidelity due to decoherence.

\subsection{Single-qubit incoherent gate error}
In modeling the incoherent error of single-qubit gates, we first consider both energy relaxation and pure dephasing. The corresponding jump operators are $\sqrt{\Gamma_1}\dyad{g}{e}$ and $\sqrt{2\Gamma_\phi}\dyad{e}{e}$, with $\Gamma_1 = 1/T_1$ and $\Gamma_\phi = 1/T_2 - 1/(2T_1)$, respectively. Here, $T_2$ is obtained from the characteristic time constant measured using the Hahn-echo sequence. By applying Eqs.\,\eqref{eq:ME_0th}–\eqref{eq:inf_ana} to each process, the incoherent error of a single-qubit gate reduces to the known form~\cite{PEDERSEN200747, DingGate}:
\begin{align}
    \bar{\epsilon}_\mathrm{1Q} := 1 - F_\mathrm{1Q} \approx \frac{1}{3}\left( \Gamma_1 + \Gamma_\phi \right) t_\mathrm{g}. \label{eq:1q_dec_err}
\end{align}

\subsubsection{Phase error due to coupler's thermal excitation \label{sec: zz_thermal_1q}}
In this study, the residual ZZ interaction between data qubits is suppressed through intrinsic static-ZZ suppression. However, not only in this work but generally in systems employing couplers, the residual ZZ interaction between a data qubit and a coupler can be significant. Consequently, thermal excitations of the coupler may induce phase errors in the data qubit via the residual coupling. To estimate the effect, we consider a probabilistic mixture determined by the thermal excitation probability. Focusing on a single-qubit X-rotation gate, the effective Hamiltonian in the rotating frame of the data qubit can be expressed depending on whether the coupler is in the ground or first excited state.
\begin{align}
    \hat{H}_\mathrm{X}/\hbar &= \frac{\Omega_\mathrm{X}}{2}\hat{X}, \\
    \hat{H}'_\mathrm{X}/\hbar &= \frac{\Omega_\mathrm{X}}{2}\hat{X} + \xi_\mathrm{ZZ}\hat{Z}.
\end{align}
Here, $\Omega_\mathrm{X}$ denotes the gate speed, which is simplified as $\Omega_\mathrm{X} = \pi/t_\mathrm{g}$ under the square pulse assumption. $\xi_\mathrm{ZZ}$ is the ZZ interaction strength between one of the data qubits and the coupler. $\hat{X}$ and $\hat{Z}$ are Pauli operators. Let $\hat{U}$ and $\hat{U}_0$ represent the ideal and actual time-evolution operators, respectively, and $\hat{M} = \hat{U}_0^\dagger\hat{U}$. The average gate fidelity is then given by~\cite{PEDERSEN200747}
\begin{align}
    F = \frac{\mathrm{Tr}[\hat{M}^\dagger \hat{M}] + |\mathrm{Tr}[\hat{M}]|^2}{d(d+1)}. \label{eq:_agf_def}
\end{align}
Expanding the average gate fidelity for the X$_\pi$ gate up to second order in $\varepsilon_\mathrm{ZZ} = \xi_\mathrm{ZZ}/\Omega_\mathrm{X}$, we obtain the approximation of the average gate fidelity as follows
\begin{align}
    F_\mathrm{ZZ} \approx 1 - \frac{8}{3}\varepsilon_\mathrm{ZZ}^2. \label{eq:1q_zz_err}
\end{align}
The thermal excitation rate of the coupler is estimated from its steady-state thermal excitation probability $P_\mathrm{th}$ and relaxation time $T_1$ as $\Gamma_{\uparrow} = P_\mathrm{th}/T_1$ from the principle of detailed balance. Assuming that the coupler's state does not change during the gate operation, the infidelity of the data qubit due to the residual ZZ interaction with the coupler can be approximated as
\begin{align}
    \bar{\epsilon}_\mathrm{ZZ} &\approx 1 - \left[ (1 - \Gamma_{\uparrow}t_\mathrm{g})\times 1 + \Gamma_{\uparrow}t_\mathrm{g} F_\mathrm{ZZ}\right] \notag\\
&\approx \frac{8}{3} \Gamma_{\uparrow}t_\mathrm{g} \varepsilon_\mathrm{ZZ}^2. \label{eq:1q_zz_err_th}
\end{align}
Then, the coherence-limited error of a single-qubit Clifford gate for each data qubit, $\bar{\epsilon}_\mathrm{1C}$, is estimated using the measured coherence times described in Sec.~\ref{sec: coherence} together with the contributions from Eqs.\,\eqref{eq:1q_dec_err} and \eqref{eq:1q_zz_err_th} as follows:
\begin{align}
    \bar{\epsilon}_\mathrm{1C} = \bar{\epsilon}_\mathrm{ZZ} + \bar{\epsilon}_\mathrm{1Q} := r_\mathrm{1C}^\mathrm{ana}. \label{eq:1Q_error_model}
\end{align}
Here, we use the ZZ interaction strengths between the coupler and the data transmons of $\xi_\mathrm{ZZ}^\mathrm{ac}=1.447(3)$\,MHz and $\xi_\mathrm{ZZ}^\mathrm{bc}=1.137(2)$\,MHz measured by the JAZZ experiments, and assume that the pulse area of a single-qubit Clifford gate is equivalent to that of the $\pi$-rotation pulse. The excitation rate of the coupler, $\Gamma_{\uparrow}$, is calculated from the steady-state excitation probability listed in Table\,\ref{tab:ro_fid}, incorporating the thermal-excitation in the $\ket{f}\mathrm{c}$ state into the thermal-excitation of the $\ket{e}\mathrm{c}$ state.

\subsubsection{Erasure-error fraction of single-qubit Clifford gates}
Since $\bar{F}_\mathrm{ZZ}$ corresponds to the error accompanied by an excitation of the coupler transmon, this error can be detected by measuring the coupler transmon. Therefore, the fraction of detectable errors for the single-qubit Clifford-gate, $R_\mathrm{1C}$, is calculated as
\begin{align}
    R_\mathrm{1C} = \frac{\bar{\epsilon}_\mathrm{ZZ}}{\bar{\epsilon}_\mathrm{1C}}. \label{eq:fraction_1Q}
\end{align}

\subsection{TIP‑based CZ-gate error}
We analyze incoherent error in the TIP‑based CZ gate. Based on Eq.\,\eqref{eq: H_fgge}, we use the following rotating‑frame Hamiltonian to implement the CZ gate:
\begin{align}
    \hat{H}_\mathrm{TIP}/\hbar &= \dyad{g}{g}_\mathrm{a}\left( -\frac{\Delta_\fgge}{4}\hat{Z}_\fgge + \frac{\Omega_\fgge}{2}\hat{X}_\fgge \right) \notag\\
    &+ \dyad{e}{e}_\mathrm{a}\left( \frac{\Delta_\fgge}{4}\hat{Z}_\fgge + \frac{\Omega_\fgge}{2}\hat{X}_\fgge \right). \label{eq:H_TIPmodel}
\end{align}
Here, the coupler transmon is modeled as a three‑level system, and the data transmons are modeled as two‑level systems. We assume that the Rabi-oscillation frequency of the $\fggetxt$ transition between Q$_\mathrm{b}$ and Q$_\mathrm{c}$ is identical across subspaces corresponding to different states of Q$_\mathrm{a}$. Using this Hamiltonian, we calculate incoherent gate errors for the decoherence processes listed in the first column of Table~\ref{tab:ana_error}. Because Hamiltonian in Eq.\,\eqref{eq:H_TIPmodel} includes states outside the computational subspace (CS), we project the density operators onto the CS with the projector $\hat{P}_\mathrm{CS}$. We then define $\hat{\rho}_0'^\psi = \hat{P}_\mathrm{CS}\hat{\rho}_0^\psi \hat{P}_\mathrm{CS}$ and $\hat{\rho}_1'^{\psi, v} = \hat{P}_\mathrm{CS}\hat{\rho}_1^{\psi, v} \hat{P}_\mathrm{CS}$, and we evaluate the state infidelity for a decoherence process $\mathcal{L}_v$ and an initial state $\ket{\psi}$ as
\begin{align}
    \bar{\epsilon}'^{\psi}_v &= 1 - \mathrm{Tr}[\hat{\rho}_0'^\psi (\hat{\rho}_0'^\psi + \hat{\rho}_1'^{\psi,v})] \notag\\
    &= - \mathrm{Tr}[\hat{\rho}_0'^\psi \hat{\rho}_1'^{\psi,v}].
    \label{eq:CZ_err_dec_phi_v}
\end{align}
Using Eq.\,\eqref{eq:CZ_err_dec_phi_v}, incoherent gate errors averaged over initial states are calculated as in Eq.\,\eqref{eq:inf_ana}. In addition, the probability of state leakage is given as
\begin{align}
    L_v^\psi = \mathrm{Tr}[\hat{P}_\mathrm{L}(\hat{\rho}_0^\psi + \hat{\rho}_1^{\psi,v})], \label{eq:leaka_cz_ana_v_psi}
\end{align}
where $\hat{P}_\mathrm{L} = \hat{I} - \hat{P}_\mathrm{CS}$ is the projector onto the leakage subspace. Note that since the leakage subspace currently under consideration includes only the first and second excited states of the coupler, all leakage can be detected through the excitation of the coupler. In Table~\ref{tab:ana_error}, the columns labeled $\bar{\epsilon}_v^\mathrm{CS}$ and $\bar{\epsilon}_v^\mathrm{L}$ report the incoherent gate errors of the TIP‑based CZ gate averaged over initial states. Here, $\bar{\epsilon}_v^\mathrm{CS}$ and $\bar{\epsilon}_v^\mathrm{L}$ denote incoherent errors in the case of $L_v^\psi=0$~(errors confined to the CS) and $L_v^\psi>0$~(leakage‑induced errors), respectively. The fourth column shows the leakage probability, $\bar{L}_v$, averaged over the initial states. From the above considerations, the total error of the TIP gate is finally given by
\begin{align}
    \bar{\epsilon}_\mathrm{CZ} = \sum_v \left(\bar{\epsilon}_v^\mathrm{CS} + \bar{\epsilon}_v^\mathrm{L} + \frac{\bar{L}_v}{4}\right) .\label{eq:err_cz_ana}
\end{align}
Using this equation, we calculate the coherence limit of the CZ-gate error. Furthermore, the last term incorporates the effect of state leakage from the CS on the gate errors~\cite{Wood_Leak}.

\begin{table}[t]
    \caption{Analytical expressions of the incoherent errors in the TIP gate.}
    \centering
    \begin{ruledtabular}
    \begin{tabular}{cccc}
    Type & $\bar{\epsilon}_v^\mathrm{CS}$ & $\bar{\epsilon}_v^\mathrm{L}$ & $\bar{L}_v$ \\ \hline
    $T_1^\mathrm{a}$ & $\Gamma_1^\mathrm{a}t_\mathrm{g}/18$        & $59\Gamma_1^\mathrm{a}t_\mathrm{g}/192$ & $9\Gamma_1^\mathrm{a}t_\mathrm{g}/128$ \\
    $T_1^\mathrm{b}$ & $2\Gamma_1^\mathrm{b}t_\mathrm{g}/9$         & $0$ & $0$ \\
    $T_1^\mathrm{c}$ & $0$           & $0$ & $0$ \\
    $T_{1f}^\mathrm{c}$ & $0$ & $3\Gamma_{1f}^\mathrm{c}t_\mathrm{g}/16$ & $3\Gamma_{1f}^\mathrm{c}t_\mathrm{g}/16$ \\
    \\
    $T_\phi^\mathrm{a}$ & $\Gamma_\phi^\mathrm{a}t_\mathrm{g}/3$          & $0$ & $0$ \\ 
    $T_\phi^\mathrm{b}$ & $0$            & $61\Gamma_\phi^\mathrm{b}t_\mathrm{g}/192$ & $21\Gamma_\phi^\mathrm{b}t_\mathrm{g}/128$ \\ 
    $T_\phi^\mathrm{c}$ & $0$            & $0$ & $0$ \\ 
    $T_{\phi f}^\mathrm{c}$ & $0$ & $15\Gamma_{\phi f}^\mathrm{c}t_\mathrm{g}/64$ & $21\Gamma_{\phi f}^\mathrm{c}t_\mathrm{g}/128$ \\ 
    \end{tabular}
    \end{ruledtabular}
    \label{tab:ana_error}
\end{table}

\subsubsection{Erasure-error fraction of the TIP‑based CZ gate}
Using the results in Table~\ref{tab:ana_error}, the fraction of detectable errors for each decoherence process can be calculated as
\begin{align}
    R_v = \frac{\bar{\epsilon}_v^\mathrm{L}}{\bar{\epsilon}_v^\mathrm{CS} + \bar{\epsilon}_v^\mathrm{L}}.
\end{align}
Note that $L_v$ is not included here. This choice ensures consistency with the gate error estimated from the decay parameter of SRB. In addition, the overall erasure-error fraction of the TIP gate is obtained as
\begin{align}
    R_\mathrm{CZ} = \frac{\sum_v \bar{\epsilon}_v^\mathrm{L}}{\sum_v ( \bar{\epsilon}_v^\mathrm{CS} + \bar{\epsilon}_v^\mathrm{L})} = \frac{\sum_v \bar{\epsilon}_v^\mathrm{L}}{r_\mathrm{CZ}^\mathrm{ana}}.
    \label{eq:tip_cz_fraction_ana}
\end{align}
Here, $r_\mathrm{CZ}^\mathrm{ana}=\sum_v ( \bar{\epsilon}_v^\mathrm{CS} + \bar{\epsilon}_v^\mathrm{L})$ denotes the analytical CZ-gate error before accounting for leakage.

\subsubsection{Error induced by transmon's charge dispersion}
The TIP gate is a two-qubit gate employing the second excited state of the coupler transmon. However, it is known that the second excited state of a transmon exhibits larger charge dispersion than the first excited state~\cite{Koch2007}. Consequently, if the frequency of the coupler's second excited state shifts due to changes in the offset charge, the operating point may deviate from the calibrated frequency, potentially leading to additional TIP-gate errors. Here, we estimate the magnitude of this effect under the assumption that the charge dispersion of the coupler's second excited state is the dominant contribution. Let the maximum frequency shift of the second excited level due to charge dispersion be $\Delta_\mathrm{c}$. In this case, Hamiltonian in Eq.\,\eqref{eq:H_TIPmodel} is modified to
\begin{align}
    \hat{H}'_\mathrm{TIP}/\hbar &= \dyad{g}{g}_\mathrm{a}\left( -\frac{\Delta_\fgge+\Delta_\mathrm{c}}{4}\hat{Z}_\fgge + \frac{\Omega_\fgge}{2}\hat{X}_\fgge \right) \notag\\
    &+ \dyad{e}{e}_\mathrm{a}\left( \frac{\Delta_\fgge-\Delta_\mathrm{c}}{4}\hat{Z}_\fgge + \frac{\Omega_\fgge}{2}\hat{X}_\fgge \right). \label{eq:H_TIPmodel_charge}
\end{align}
Let $\hat{U}$ and $\hat{U}_0$ denote the time-evolution operators under Hamiltonians~\eqref{eq:H_TIPmodel} and\,\eqref{eq:H_TIPmodel_charge}, respectively. From Eq.\,\eqref{eq:_agf_def}, the average gate fidelity including the effect of charge dispersion, $F_\mathrm{CZ}^\mathrm{c}$, can be approximated as
\begin{align}
    F_\mathrm{CZ}^c &= \frac{\mathrm{Tr}[\hat{M}'^\dagger \hat{M}'] + |\mathrm{Tr}[\hat{M}']|^2}{20} \notag\\
    &\approx 1 - \frac{\pi^2}{640}\varepsilon_\mathrm{c}^2 \label{eq:cz_err_charge_disp}
\end{align}
where $\hat{M}' = \hat{P}_\mathrm{CS}\hat{U}_0^\dagger \hat{U} \hat{P}_\mathrm{CS}$, and the final expression corresponds to a second-order expansion in $\varepsilon_\mathrm{c} = \Delta_\mathrm{c} / \Delta_\fgge$. The maximum charge dispersion of the second excited state of the transmon is given by~\cite{Koch2007}
\begin{align}
    \hbar\Delta_\mathrm{c} = (-1)^2E_\mathrm{C}\frac{2^{13}}{2!}\sqrt{\frac{2}{\pi}}\left( \frac{E_\mathrm{J}}{2E_\mathrm{C}} \right)^{\frac{7}{4}}e^{-\sqrt{8E_\mathrm{J}/E_\mathrm{C}}}.
\end{align}
Using the values listed in Table~\ref{tab:device_params}, we obtain $E_\mathrm{J}/E_\mathrm{C}\sim59$ and $E_\mathrm{C}/\hbar\sim2\pi\times300$~MHz, and the estimated frequency shift due to the charge dispersion is $\Delta_\mathrm{c}/2\pi \sim 200$~kHz. From Eq.\,\eqref{eq:cz_err_charge_disp} and $\Delta_\fgge/2\pi\sim10$~MHz of the present device, the resulting gate infidelity is on the order of $10^{-4}$ or less. Moreover, this contribution is further suppressed by the product of the reported offset-charge tunneling rate of about $1$~kHz~\cite{ChargeP_2013, ChargeP_2019} and the gate time of $140$~ns. Thus, we conclude that the overall effect of offset-charge noise on the TIP gate is negligible under the present device parameter regime. Therefore, this effect is not included in the analysis of the coherence limit or the erasure-error fraction.

\section{\label{sec: fraction_cal}Derivation of Eq.~(18)}
\begin{figure}
    \centering
    \includegraphics{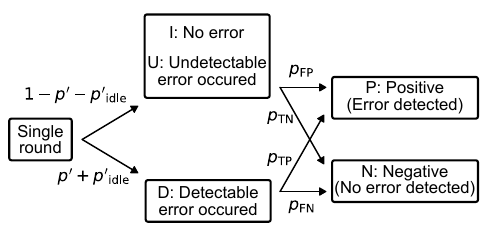}
    \caption{Error processes in a single round of the SRB(MCM). The probability $p' + p'\mathrm{idle}$ represents the occurrence of a detectable error. The events I, U, and D correspond to cases in which no error, an undetectable error, and a detectable error actually occurred, respectively. P and N denote whether an error is detected or not detected, respectively. The probabilities $p_\mathrm{TN}$ and $p_\mathrm{FN}$ represent correctly and incorrectly identifying an event as U or I, respectively, while $p_\mathrm{TP}$ and $p_\mathrm{FP}$ represent correctly and incorrectly identifying an event as D.}
    \label{fig:error_prop}
\end{figure}

Here we explain the derivation of Eq.\,\eqref{eq:frac}.
The error of a two-qubit Clifford gate is defined as
\begin{align}
    r_\mathrm{2C} := p + p'. \label{eq: J_r2C}
\end{align}
Here, the probabilities $p$ and $p'$ represent undetectable and detectable errors by the MCM of the coupler, respectively. The left-hand side of the equation corresponds to the value experimentally estimated from the SRB results. The erasure error fraction $R$ is defined as \cite{Wu2022}
\begin{align}
    R \equiv \frac{\text{Detected error}}{\text{Total error}}.
\end{align}
Next, we explain how $R$ is estimated from the experimental data.
When the SRB(MCM) data are fitted to an exponential decay without the post-selection, the obtained gate error $r_\mathrm{w/o}$ can be expressed as
\begin{align}
    r_\mathrm{w/o} = r_\mathrm{2C} + r_\mathrm{idle} = p + p' + p_\mathrm{idle} + p'_\mathrm{idle}, \label{eq: J_rwo}
\end{align}
where we assume that errors from the two-qubit Clifford gate and from the idling do not occur simultaneously. The additional idle error, $r_\mathrm{idle}$, introduced by the MCM pulse can also be separated into detectable ($p_\mathrm{idle}$) and undetectable ($p'_\mathrm{idle}$) error probabilities. When the post-selection is applied to the SRB(MCM) data and only the results where the MCM outcome remains in the ground state are used, the obtained error probability $r_\mathrm{w/}$ is expressed as
\begin{align}
    r_\mathrm{w/} &= P(\mathrm{U}|\mathrm{N}) + P(\mathrm{D}|\mathrm{N}) \notag\\
    &= \frac{P(\mathrm{N}|\mathrm{U})P(\mathrm{U}) + P(\mathrm{N}|\mathrm{D})P(\mathrm{D})}{P(\mathrm{N})} \notag\\
    &= \frac{p_\mathrm{TN}(p + p_\mathrm{idle}) + p_\mathrm{FN}(p'+p'_\mathrm{idle})}{p_\mathrm{TN}(1 - p' - p'_\mathrm{idle}) + p_\mathrm{FN}(p'+p'_\mathrm{idle})}. \label{eq: J_rw}
\end{align}
Here, $P(\mathrm{A}|\mathrm{B})$ denotes the conditional probability that event $\mathrm{A}$ occurs given event $\mathrm{B}$, and undetectable ($\mathrm{U}$) and detectable ($\mathrm{D}$) events are assumed to be mutually exclusive. The quantities $p_\mathrm{TN}$ and $p_\mathrm{FN}$ represent the probabilities of correct and incorrect identifications of a detectable error, respectively (see Fig.\,\ref{fig:error_prop} for details). By solving Eqs.\,\eqref{eq: J_rw} and \eqref{eq: J_rwo} for $p$ and $p'$, and assuming $1 \gg \{p, p', p_\mathrm{idle}, p_\mathrm{FN}\}$, we obtain the first-order expression of the erasure error fraction as 
\begin{align}
    R = \frac{p'}{p + p'} \approx \frac{\Delta r_\mathrm{PED} - p'_\mathrm{idle}}{r_\mathrm{2C}}=:R_\mathrm{1st}.
\end{align}

\subsection{Estimation of the detectable idling error}
We estimate the magnitude of phase errors arising from the coupler excitation from $\ket{g}_\mathrm{c}$ to $\ket{e}_\mathrm{c}$ to be the dominant source of the detectable error $p'_\mathrm{idle}$ during the idling introduced by the MCM pulse. The ideal time evolution during idling is the identity, and the error Hamiltonian is
\begin{align}
    \hat{H}_\mathrm{idle}/\hbar = \xi_\mathrm{ZZ}^\mathrm{ac}\hat{Z}_\mathrm{a} + \xi_\mathrm{ZZ}^\mathrm{bc}\hat{Z}_\mathrm{b}.
\end{align}
From Eq.\,\eqref{eq:_agf_def}, the infidelity between the evolution governed by the above Hamiltonian and the identity at time $t$ is
\begin{align}
\epsilon_\mathrm{idle}'(t) :&= 1 - F_\mathrm{idle}'(t) \notag\\
&= \frac{4}{5}\left[1-\cos^2(\xi_\mathrm{ZZ}^\mathrm{a}t)\cos^2(\xi_\mathrm{ZZ}^\mathrm{b}t) \right]. 
\end{align}
Because an excitation occurring in the first half of the readout-window duration $t_\mathrm{m}=420$~ns is detectable, the detectable error during the idling can be approximated, similar to the single-qubit Clifford-gate case, as
\begin{align}
    p'_\mathrm{idle} \approx \Gamma_{\uparrow}\left(\frac{t_\mathrm{m}}{2} + t_\mathrm{pp}\right)\epsilon_\mathrm{idle}'(t_\mathrm{m}),
\end{align}
where $t_\mathrm{pp}=10$\,ns is the buffer time between the two-qubit Clifford gate and the MCM pulse. As in Sec.~\ref{sec: zz_thermal_1q}, using the experimental ZZ interaction strengths between the coupler and the data transmons and the excitation rate of the coupler, we estimate the detectable error during idling to be $p'_\mathrm{idle} \approx0.056(8)\%$. As mentioned in Sec.~\ref{sec: leak_suv_ped}, the coupler’s excitation and relaxation rates during the readout can differ from their original values. The effect of the dispersive readout on the transition rates is asymmetric between the excited and ground states, and the ground state is weakly affected \cite{MIST_disp}. The observation that there is no increase in leakage from the computational subspace when the MCM pulses are applied supports this behavior. Consequently, we use the excitation rate $\Gamma_{\uparrow}$ calculated from the $T_1$ and the thermal excitation probability $P_\mathrm{th}$ of the coupler.

\bibliography{references}

\end{document}